\documentclass[fleqn,usenatbib]{mnras}
\usepackage[T1]{fontenc}
\usepackage{ae,aecompl}
\usepackage[dvipdfmx]{graphicx}
\usepackage{amsmath}	
\usepackage{amssymb}	
\usepackage{lscape}
\usepackage{graphicx}
\usepackage{bm}

\title[Intermittent accretion in SMS formation]{Disk fragmentation and intermittent accretion onto supermassive stars}

\author[R. Matsukoba et al.]{
Ryoki Matsukoba,$^{1}$\thanks{E-mail: r.matsukoba@astr.tohoku.ac.jp}
Eduard I. Vorobyov,$^{2,3}$
Kazuyuki Sugimura,$^{1,4}$ 
\newauthor
~Sunmyon Chon,$^{1}$ Takashi Hosokawa,$^{5}$ and 
Kazuyuki Omukai$^{1}$
\\
$^{1}$Astronomical Institute, Graduate School of Science, Tohoku University, Aoba, Sendai, Miyagi 980-8578, Japan\\
$^{2}$University of Vienna, Department of Astrophysics, Vienna, 1180, Austria\\
$^{3}$Ural Federal University, 51 Lenin Str., 620051 Ekaterinburg, Russia\\
$^{4}$Department of Astronomy, University of Maryland, College Park, MD, 20740, US\\
$^{5}$Department of Physics, Graduate School of Science, Kyoto University, Sakyo, Kyoto 606-8502, Japan
}
\date{Accepted XXX. Received YYY; in original form ZZZ}

\pubyear{2020}

\begin{document}
\label{firstpage}
\pagerange{\pageref{firstpage}--\pageref{lastpage}}
\maketitle

\begin{abstract}
Supermassive stars (SMSs) with $\sim10^{4-5}~\mathrm{M}_{\odot}$ are candidate objects 
for the origin of supermassive black holes observed at redshift $z$>6. 
They are supposed to form in primordial-gas clouds that provide the central stars with gas at a high accretion rate, 
but their growth may be terminated in the middle due to the stellar ionizing radiation 
if the accretion is intermittent and its quiescent periods are longer than the Kelvin-Helmholtz (KH) timescales at the stellar surfaces. 
In this paper, we examine the role of the ionizing radiation feedback 
based on the accretion history in two possible SMS-forming clouds extracted from cosmological simulations, 
following their evolution with vertically-integrated two-dimensional hydrodynamic simulations with detailed thermal and chemical models. 
The consistent treatment of the gas thermal evolution is crucial for obtaining the realistic accretion history, 
as we demonstrate by performing an additional run with a barotropic equation of state, 
in which the fluctuation of the accretion rate is artificially suppressed. 
We find that although the accretion becomes intermittent due to the formation of spiral arms and clumps in gravitationally unstable disks, 
the quiescent periods are always shorter than the KH timescales, 
implying that SMSs can form without affected by the ionizing radiation. 
\end{abstract}



\section{Introduction}
\label{Sec:1}

More than 200 supermassive black holes (SMBHs) with 10$^7$-10$^{10}$~$\mathrm{M}_{\odot}$ at redshift $z>6$ have been discovered 
by recent observations of high-redshift quasars 
(e.g. \citealt{Venemans:2013}; \citealt{Banados:2018}; \citealt{Matsuoka:2018}; \citealt{Onoue:2019}; see also \citealt{Gallerani:2017} for a review). 
Although the standard formation scenario explaining the origin of these BHs has not yet been established, 
massive seed BHs are preferred because the existence of the high-redshift SMBHs suggests that they have to grow to SMBHs in a short time 
(see, e.g., \citealt{Volonteri:2012}; \citealt{Haiman:2013}; \citealt{Inayoshi:2019} for a review).

Remnant BHs of Pop III stars have been considered as candidates of seed BHs by some authors (e.g., \citealt{Madau:2001}). 
They possibly grow to the observed high-redshift SMBHs, 
either by continuous accretion at Eddington limit or by short episodic accretion at a super-Eddington rate.
In practice, however, it is hard to realize such accretion growths,
because accretion flows onto seed BHs are easily inhibited by their own radiation together with the gas angular momentum
(\citealt{Milosavljevic:2009}; \citealt{Park:2011}; \citealt{Sugimura:2018}; but see also \citealt{Inayoshi:2016}; \citealt{Sugimura:2017}).

An alternative SMBH formation channel is the so-called direct collapse scenario (e.g., \citealt{Bromm:2003}), 
in which supermassive stars (SMSs) with $\sim10^{4-5}~\mathrm{M}_{\odot}$ collapse into seed BHs with the similar mass after their lifetime \citep{Umeda:2016}. 
The SMSs are supposed to form in primordial-gas clouds if 
the clouds collapse almost isothermally at $\sim10^4$ K due to the atomic hydrogen cooling,
with the formation of molecular hydrogen fully suppressed by strong external far-ultraviolet (UV) radiation from nearby galaxies 
\citep{Omukai:2001, Shang:2010, Regan:2014, Sugimura:2014}. 
The high gas temperature of the clouds leads to a
high accretion rate of $0.1-1~\mathrm{M}_{\odot}~\mathrm{yr}^{-1}$ 
onto the protostars formed at the center \citep{Latif:2013, Inayoshi:2014-11, Becerra:2015},
as well as prevents the vigorous gas fragmentation in the clouds. 
Due to the high accretion rate, 
the surface of protostars substantially inflates and the effective temperature drops to several 1000 K \citep{Hosokawa:2012, Hosokawa:2013}.
As a result, the accretion flow continues without affected by the radiative feedback,
allowing the protostars to reach the mass
of $\sim10^{4-5}~\mathrm{M}_{\odot}$ within their short lifetime ($\sim$Myr).

In order to maintain the inflated stellar surface
with a constant accretion rate,
the accretion rate must be higher than the critical value of $4\times10^{-2}~\mathrm{M}_{\odot}~\mathrm{yr}^{-1}$,
as shown in \cite{Omukai:2003, Hosokawa:2012, Hosokawa:2013} 
(but see \citealt{Haemmerle:2018-2} for discussion that the critical value decreases below 10$^{-2}$~M$_{\odot}$~yr$^{-1}$ 
if the stellar mass is above 600~M$_{\odot}$). 
If the accretion rate temporarily drops below the critical value for sometime, 
the stellar surface begins to shrink, and hence 
the effective temperature rises.
The ionizing radiation from the shrinking protostar may quench the accretion
before acquiring enough mass to reach the SMS regime. 
By performing stellar evolution calculations with an accretion model with repeating burst and quiescent phases, 
\cite{Sakurai:2015} showed that 
ionizing radiation from the protostar becomes strong enough to significantly suppress the accretion
if a quiescent period of the intermittent accretion $\Delta t_{\mathrm{q}}$, 
which is defined as the time duration for which 
the accretion rate is below the critical value,
is longer than the Kelvin-Helmholtz (KH) timescale at the stellar surface
\begin{align}
t_{\mathrm{KH,surf}} = 10^{3}~\mathrm{yr}~\left( \frac{M_{\ast}}{500~\mathrm{M_{\odot}}} \right)^{1/2}~.
\label{Eq:1}
\end{align}

The time variation of the accretion rate can be caused by
the fragmentation of the circumstellar disks due to the gravitational instability,
as suggested in the stability analysis of the disks around growing SMSs \citep{Inayoshi:2014-12, Latif:2015, Matsukoba:2019}.
\cite{Sakurai:2016} confirmed that the disk fragmentation due to the gravitational instability in fact
causes the fluctuation of the accretion rate,
by performing vertically-integrated two-dimensional simulations of the disks around growing SMSs. 
From the stellar evolution calculations with the accretion rate obtained from the simulations,
they also concluded that SMSs can grow by accretion without affected by the radiative feedback.
Consistently, the quiescent periods observed in their simulations were always shorter than the KH timescale given in Equation~\eqref{Eq:1}.

Their simulations, however, adopted a barotropic equation of state to model the thermal evolution of gas,
instead of solving the energy equation. 
Considering that the temperature of the gas plays a critical role in determining the
gravitational stability of the disks, this approximation may affect their conclusion on the role of radiative feedback.
Most importantly, their barotropic relation was an inadequate approximation 
because it is based on the thermal evolution of collapsing cores \citep{Omukai:2001}, 
which is largely different from that of disks \citep{Matsukoba:2019}.
Therefore, in this paper, 
we perform simulations of SMS formation considering detailed thermal and chemical processes
and re-examine whether the protostars can grow to SMSs without affected by the radiative feedback.

The paper is organized as follows. 
We describe our simulation model and the initial conditions in Section~\ref{Sec:2}.
We then present the time evolution of disk structures and central stars, as well as
the comparison with a simulation with a barotropic relation, in Section~\ref{Sec:3}. 
Summary and discussion are given in Section~\ref{Sec:4}.

\section{Method}
\label{Sec:2}

We follow the time evolution of the disks around growing SMSs
using vertically-integrated two-dimensional simulations with a detailed treatment of chemical and thermal processes.
Here, we first briefly explain the method for the hydrodynamic simulations
and then describe the thermal processes, chemical reactions, and initial conditions adopted in this study. 
The details of the hydrodynamic method are described in \cite{Vorobyov:2020-6}.

\subsection{Hydrodynamic simulations}
\label{Sec:2-1}

Here, we describe the method for our vertically-integrated two-dimensional simulations used to follow the gas dynamics around growing SMSs.
We use polar-coordinate ($r,~\phi$) grids with 512$\times$512 spatial zones. 
The computational domain extends to the outer radius of  $r_{\mathrm{out}}=2\times10^6$ au, with
the sink cell with the size $r_{\mathrm{sc}}=300$ au introduced at the center. 
At each time step, we measure the mass flowing into the sink cell, 
in which we assume that a central star is surrounded by an unresolved disk,
and increase the stellar mass according to the following sink-cell model:
4 \% of the gas flowing into the sink cell is deposited in the unresolved disk,
9.6 \% is carried away by the stellar jet, and the rest accretes onto the central star. 
We initially set the stellar mass to zero and 
the surface density of the sink cell to the same as in the innermost grids. 

To follow the hydrodynamic evolution of the gas, we solve the vertically-integrated
mass, momentum, and energy transport equations:
\begin{align}
&\frac{\partial\Sigma}{\partial t} = -\nabla_{\mathrm{p}}\cdot\left( \Sigma\bm{u}_{\mathrm{p}} \right)~, \label{Eq:2} \\
&\frac{\partial}{\partial t}\left( \Sigma\bm{u}_{\mathrm{p}} \right) 
+ \left[ \nabla\cdot \left( \Sigma\bm{u}_{\mathrm{p}}\otimes\bm{u}_{\mathrm{p}} \right)  \right]_{\mathrm{p}}
= -\nabla_{\mathrm{p}}P + \Sigma\bm{g}_{\mathrm{p}} + \left( \nabla\cdot\bm{\Pi} \right)_{\mathrm{p}}~, \label{Eq:3} \\
&\frac{\partial e}{\partial t} + \nabla_{\mathrm{p}}\cdot\left( e\bm{u}_{\mathrm{p}} \right)
= -P\left( \nabla_{\mathrm{p}}\cdot\bm{u}_{\mathrm{p}} \right) - Q_{\mathrm{net}} 
+ \left( \nabla \bm{u} \right)_{\mathrm{pp'}}:\bm{\Pi}_{\mathrm{pp'}}~, \label{Eq:4}
\end{align}
where the subscripts $\mathrm{p}$ and $\mathrm{p}'$ represents the planar components ($r,~\phi$) in the polar coordinates, 
$\Sigma$ is the surface density, 
$\bm{u}_{\mathrm{p}}=u_{\mathrm{r}}\bm{\hat{r}}+u_{\mathrm{\phi}}\bm{\hat{\phi}}$ is the planar velocity, 
$P$ is the vertically integrated gas pressure, 
$\nabla=\bm{\hat{r}}\partial/\partial r+\bm{\hat{\phi}}r^{-1}\partial/\partial\phi$ is the gradient in the disk plane, 
$\bm{g}_{\mathrm{p}}=g_{\mathrm{r}}\bm{\hat{r}}+g_{\mathrm{\phi}}\bm{\hat{\phi}}$ is the gravitational acceleration including the gravity of the central star and the self-gravity of the circumstellar disk, 
$e$ is the internal energy per unit area, and 
$Q_{\mathrm{net}}$ is the net cooling rate per unit area, which we describe in Section~\ref{Sec:2-2}. 
The gas pressure and internal energy are 
related by the ideal-gas equation of state, 
\begin{align}
P = (\gamma-1)e~,
\label{Eq:5}
\end{align}
with the adiabatic exponent $\gamma$, which we consistently calculate according to the chemical composition
considering the rotational and vibrational degrees of freedom of the H$_2$. 
The gas mass density and temperature, which are used for the computation of
the thermal and chemical evolution,
are given respectively by
\begin{align}
\rho = \frac{\Sigma}{\sqrt{2\pi}\,H_{\mathrm{g}}} 
\label{Eq:6}
\end{align}
and
\begin{align}
T = (\gamma-1) \frac{\mu m_{\mathrm{H}}}{k_{\mathrm{B}}}\frac{e}{\Sigma}~,
\label{Eq:7}
\end{align}
where
$H_{\mathrm{g}}$ is the gas scale height 
estimated from vertical hydrostatic balance in the gravitational fields of the star and disk (see \citealt{Vorobyov:2009-3}), 
$\mu$ is the mean molecular weight, $k_{\mathrm{B}}$ is the Boltzmann constant, 
and $m_{\mathrm{H}}$ is the mass of a hydrogen nucleus.
The self-gravity of the disk is computed by taking the gradient of the gravitational potential 
\begin{align}
\Phi(r,~\phi) &= -G\int^{r_{\mathrm{out}}}_{r_{\mathrm{sc}}} r'\mathrm{d}r' \notag\\
&\times \int^{2\pi}_{0} \frac{\Sigma(r',~\phi')}{\sqrt{r'^{2}+r^{2}-2rr'\mathrm{cos}(\phi'-\phi)}}~\mathrm{d}\phi'~.
\label{Eq:8}
\end{align}
The turbulent viscosity is considered with the viscous stress tensor 
\begin{align}
\bm{\Pi} = 2\Sigma\nu\left( \nabla\bm{u}-\frac{1}{3}(\nabla\cdot\bm{u})\bm{\mathrm{e}} \right)~,
\label{Eq:9}
\end{align}
where $\bm{\mathrm{e}}$ is the unit tensor and 
$\nu$ is the kinematic viscosity, which 
is given according to the $\alpha$-viscosity prescription \citep{Shakura:1973}, 
\begin{align}
\nu = \alpha c_{\mathrm{s}}H_{\mathrm{g}}~.
\label{Eq:10}
\end{align}
Here, $c_{\mathrm{s}}=\sqrt{\gamma P / \Sigma}$ is the sound velocity. 
In this study, we set $\alpha=10^{-4}$. 
Although we consider the angular momentum transport due to
the turbulent viscosity, the primary angular momentum transport mechanism is that due to the gravitational torque.

\subsection{Thermal processes}
\label{Sec:2-2}

The net cooling rate per unit area is given by 
\begin{align}
Q_{\mathrm{net}} = \int \Lambda_{\mathrm{net}}~\mathrm{d}z = 2H_{\mathrm{g}}\Lambda_{\mathrm{net}}~,
\label{Eq:11}
\end{align}
where $\Lambda_{\mathrm{net}}$ is the net cooling rate per unit volume.
The value of $\Lambda_{\mathrm{net}}$ is the sum of the rates of H$_{2}$-line cooling $\Lambda_{\mathrm{H_{2}}}$, Lyman-$\alpha$ cooling $\Lambda_{\mathrm{Ly\alpha}}$, 
continuum cooling $\Lambda_{\mathrm{cont}}$, chemical cooling $\Lambda_{\mathrm{chem}}$, 
H$^{-}$ photodetachment heating $\Gamma_{\mathrm{PD}}$, and stellar irradiation heating $\Gamma_{\mathrm{irr}}$:
\begin{align}
\Lambda_{\mathrm{net}} =  \Lambda_{\mathrm{H_{2}}} + \Lambda_{\mathrm{cont}} + \Lambda_{\mathrm{Ly\alpha}} 
+ \Lambda_{\mathrm{chem}} - \Gamma_{\mathrm{PD}} - \Gamma_{\mathrm{irr}}~.
\label{Eq:12}
\end{align}

The H$_{2}$-line cooling rate is given by
\begin{align}
\Lambda_{\mathrm{H}_{2}} = \overline{\beta}_{\mathrm{esc,H_{2}}}\Lambda_{\mathrm{H}_{2},\mathrm{thin}}\mathrm{e}^{-\tau}~,
\label{Eq:13}
\end{align}
where $\Lambda_{\mathrm{H_{2},thin}}$ is the optically-thin rate \citep{Glover:2015-11},
$\overline{\beta}_{\mathrm{esc,H_{2}}}$ is the line-averaged escape probability  \citep{Fukushima:2018},
and $\tau$ is the effective optical depth for continuum radiation. 
The effective optical depth 
\begin{align}
\tau = \sqrt{\tau_{\mathrm{P}}\tau_{\mathrm{R}}}~, 
\label{Eq:14}
\end{align}
is calculated with the Planck (Rosseland) mean optical depth 
\begin{align}
\tau_{\mathrm{P(R)}} = \frac{1}{2}\Sigma\,\kappa_{\mathrm{P(R)}}~,
\label{Eq:15}
\end{align}
for which we use the Planck (Rosseland) mean opacity $\kappa_{\mathrm{P(R)}}$ provided by \cite{Mayer:2005}.
Similarly, the Lyman-$\alpha$ cooling rate is given by
\begin{align}
\Lambda_{\mathrm{Ly}\alpha} = \overline{\beta}_{\mathrm{esc,Ly\alpha}}\Lambda_{\mathrm{Ly}\alpha,\mathrm{thin}}\mathrm{e}^{-\tau}~,
\label{Eq:16}
\end{align}
where $\Lambda_{\mathrm{Ly}\alpha,\mathrm{thin}}$ is optically-thin rate \citep{Cen:1992} 
and $\overline{\beta}_{ \mathrm{esc,Ly\alpha}}$ is the escape probability estimated by using the method in \cite{Inayoshi:2016}.
We consider H free-bound emission, H$^{-}$ free-bound emission, H$^{-}$ free-free emission, 
H free-free emission, H$_{2}$-H$_{2}$ collision-induced emission, and H$_{2}$-He collision-induced emission as the continuum radiation processes. 
The H$^{-}$ free-bound emission plays the primary role as a coolant in the circumstellar disks \citep{Matsukoba:2019}. 
We use the fitting formula for the continuum cooling rate in the optically thin regime $\Lambda_{\mathrm{cont,thin}}$ from \cite{Matsukoba:2019} 
and smoothly connect the rates in the optically thin and thick limits \citep{Becerra:2018}:
\begin{align}
\Lambda_{\mathrm{cont}} = \Lambda_{\mathrm{cont},\mathrm{thin}}\left(1+\frac{3}{2}\tau^{2}\right)^{-1}~.
\label{Eq:17}
\end{align}
The chemical cooling/heating processes include H ionization/recombination, H$_{2}$ dissociation/formation, and 
H$^{-}$ detachment/attachment. 
The chemical cooling rate is calculated as follows:
\begin{align}
\Lambda_{\mathrm{chem}} = \left( \frac{\mathrm{d}y(\mathrm{H}^{+})}{\mathrm{d}t}\chi_{\mathrm{H}} 
- \frac{\mathrm{d}y(\mathrm{H}_{2})}{\mathrm{d}t}\chi_{\mathrm{H}_{2}} 
- \frac{\mathrm{d}y(\mathrm{H}^{-})}{\mathrm{d}t}\chi_{\mathrm{H}^{-}} \right) n_{\mathrm{H}}~,
\label{Eq:18}
\end{align}
where $\chi_{\mathrm{H}}$=13.6 eV, $\chi_{\mathrm{H}_{2}}$=4.48 eV, and $\chi_{\mathrm{H}^{-}}$=0.755 eV are the binding energies.
The chemical fraction of species $i$, $y(i)$, is defined by the ratio of its number density $n(i)$ and that of hydrogen nuclei $n_{\mathrm{H}}$:
\begin{align}
y(i) = \frac{n(i)}{n_{\mathrm{H}}}~.
\label{Eq:19}
\end{align}
The number density of hydrogen nuclei is given by 
\begin{align}
n_{\mathrm{H}} = \frac{\rho}{\left(1+4y_{\mathrm{He}}\right)m_{\mathrm{H}}}~,
\label{Eq:20}
\end{align}
where $y_{\mathrm{He}}$ is the fractional abundance of helium.

In SMS formation, 
H$^{-}$ photodetachment by external radiation may contribute to the suppression of H$_{2}$ formation in the low-density region. 
The gas is also heated upon photodetachment because the excess photon energy is stored as the kinetic energy of photodetached free electrons. 
The H$^{-}$ photodetachment heating rate is given by 
\begin{align}
\Gamma_{\mathrm{PD}} = \epsilon_{\mathrm{PD}}\,n_{\mathrm{H}}\,y({\mathrm{H}^-})\,k_{22}~,
\label{Eq:21}
\end{align}
where $\epsilon_{\mathrm{PD}}$ is the average heating rate per reaction,
$k_{22}$ is the photodetachment rate per $\mathrm{H}^-$ ion (reaction number 22 in Table~\ref{Tab:a1}). 
The average heating rate per reaction is calculated as 
\begin{align}
\epsilon_{\mathrm{PD}} = \frac{\int 4\pi\frac{J_{\mathrm{ex}}(\nu)}{h\nu}\sigma_{\mathrm{PD}}(\nu)h\nu~\mathrm{d}\nu}
{\int 4\pi\frac{J_{\mathrm{ex}}(\nu)}{h\nu}\sigma_{\mathrm{PD}}(\nu)~\mathrm{d}\nu}~,
\label{Eq:22}
\end{align}
with the reaction cross-section $\sigma_{\mathrm{PD}}$ \citep{John:1988} and 
the external radiation intensity $J_{\mathrm{ex}}(\nu)$.
As in \cite{Chon:2018}, we simply assume the blackbody radiation spectrum
\begin{align}
J_{\mathrm{ex}}(\nu) &= 10^{-21}J_{21} \notag \\
&\times \frac{B_{\nu}(T_{\mathrm{ex}})}{B_{13.6~\mathrm{eV}}(T_{\mathrm{ex}})}\mathrm{e}^{-\tau}~\mathrm{erg~s^{-1}~Hz^{-1}~str^{-1}~cm^{-2}}~,
\label{Eq:23}
\end{align}
with the Planck function  $B_{\nu}(T_{\mathrm{ex}})$ 
and the far-UV intensity $J_{21}$  (in the unit of $10^{-21}~\mathrm{erg~s^{-1}~Hz^{-1}~str^{-1}~cm^{-2}}$ at $h\nu=13.6$ eV), 
and set the radiation temperature $T_{\mathrm{ex}}=10^{4}$ K 
\citep[but see also][for discussion about realistic radiation spectra]{Sugimura:2014}.
This yields $\epsilon_{\mathrm{PD}}=2.23$ eV, independently of $\tau$ and $J_{21}$, 

Our thermal model also takes into account the central stellar irradiation heating. 
As the star grows and its luminosity increases, 
it may affect the gas temperature. 
The stellar irradiation heating rate is calculated as 
\begin{align}
\Gamma_{\mathrm{irr}} = \frac{4\sigma_{\mathrm{SB}}\rho}{1+\frac{3}{2}\tau^{2}} \kappa_{\mathrm{P}}(T_{\mathrm{irr}})T_{\mathrm{irr}}^{4}~,
\label{Eq:24}
\end{align}
with the Stefan-Boltzmann constant $\sigma_{\mathrm{SB}}$ and the irradiation temperature $T_{\mathrm{irr}}$,
which is given by 
\begin{align}
T_{\mathrm{irr}} = \left( G(\tau)\frac{L_{\ast}}{4\pi\sigma_{\mathrm{SB}}r^{2}} \right)^{1/4}~.
\label{Eq:25}
\end{align}
The function $G(\tau)$ smoothly connects the values in both the optically thin and thick regimes:
\begin{align}
G(\tau) = \frac{1}{4} + \frac{2}{\pi} \left( \mathrm{cos}\gamma_{\mathrm{irr}}-\frac{1}{4} \right) \mathrm{arctan(\tau)}~,
\label{Eq:26}
\end{align}
with the incident angle of stellar irradiation to the disk $\gamma_{\mathrm{irr}}$ \citep{Vorobyov:2010}.
This function becomes 1/4 in the optically thin regime and cos$\gamma_{\mathrm{irr}}$ in the optically thick regime. 
We compute the stellar luminosity $L_{\ast}$ 
using the analytical formula obtained from stellar evolution calculations \citep{Hosokawa:2012}: 
\begin{align}
L_{\ast} = 3.8\times10^{6}~\mathrm{L}_{\odot}~\left( \frac{M_{\ast}}{100~\mathrm{M}_{\odot}} \right)~,
\label{Eq:27}
\end{align}
where $M_{\ast}$ is the stellar mass.

It is known that artificial fragmentation occurs in hydrodynamic simulations 
if the Jeans length $\lambda_{\mathrm{J}}$ becomes less than four times the grid size $x_{\mathrm{grid}}$ \citep{Truelove:1997}.
In order to prevent such artificial fragmentation, we cut off the cooling by 
introducing a suppression factor \citep{Hosokawa:2016},
\begin{align}
&C_{\mathrm{limit}} = \mathrm{exp}\left[ -\left( \frac{\xi-1}{0.1} \right)^{2} \right]~,
\label{Eq:28} \\
&\xi = f_{\mathrm{limit}}\frac{x_{\mathrm{grid}}}{\lambda_{\mathrm{J}}}~,
\label{Eq:29}
\end{align}
and multiplying $\Lambda_{\mathrm{net}}$ by this factor.
We set $f_{\mathrm{limit}}=6$ in our model, 
and hence the cooling is suppressed when $\lambda_{\mathrm{J}}$ becomes less than six times $x_{\mathrm{grid}}$.

For comparison with the previous study \citep{Sakurai:2016},
we also perform hydrodynamic simulations using the barotropic temperature-density relation described in Appendix~\ref{App:b}, 
instead of solving the energy equation (Equation~\ref{Eq:4}).
We describe the results from the simulations with the barotropic relation
in Section~\ref{Sec:3-3}.

\subsection{Chemical reactions}
\label{Sec:2-3}

We follow the chemical evolution of the primordial gas,
solving the chemical network of five species, H, H$_{2}$, H$^{+}$, H$^{-}$, and e,
and 22 reactions, summarized in Table~\ref{Tab:a1}. 
Our chemical network was selected so as to correctly follow the thermal evolution of
both collapsing clouds and circumstellar disks in the SMS formation \citep{Omukai:2001,Matsukoba:2019}. 
In our chemical model,
we solve the non-equilibrium kinetic equations for H, H$_{2}$, H$^{+}$, and e, 
with H$^{-}$ assumed to be in the chemical equilibrium of all related reactions. 
We assume that all helium is neutral, with the fractional abundance $y_{\mathrm{He}}=8.333\times10^{-2}$. 
We further solve the continuity equation for each species assuming the collisional coupling with the gas.


\subsection{Initial conditions}
\label{Sec:2-4}

We start our simulations from the initial conditions extracted from the previous cosmological simulations in \cite{Chon:2016} 
and follow the SMS formation from the pre-stellar core stage until the masses of the central stars reach 30000~M$_{\odot}$. 
\cite{Chon:2016} performed tens of zoom-in hydrodynamic simulations in a parent volume of 30 Mpc on a side and identified
two collapsing primordial gas clouds that are exposed by strong far-UV radiation from nearby galaxies and possibly form SMSs later on.
We extract these two clouds, which were labelled {\it filamentary} and {\it spherical} clouds from their shapes,
when the core density reaches 10$^5$ cm$^{-3}$.

\begin{table*}
 \begin{center}
 \caption{Initial properties of the simulated clouds}
 \label{Tab:1}
  \scalebox{0.98}[0.98]{ 
  \begin{tabular}{l c c c c c c c c} \hline \hline
     & $r_{\mathrm{c}}$ (pc) & $M_{\mathrm{c}}$ (M$_{\odot}$) & $M_{\mathrm{tot}}$ (M$_{\odot}$) & $\Omega_{\mathrm{c}}$ (s$^{-1}$) & $J_{21}$ 
     & $T$ (K) & $y$(H$_{2}$) & $y$(H$^{+}$), $y$(e) \\ \hline
     {\it filamentary} & 1.35 & $5.5\times10^{4}$ & $7.9\times10^{5}$ & $5.8\times10^{-14}$ & 5000  &7100 & $1.8\times10^{-9}$ & $7.3\times10^{-5}$ \\
     {\it spherical}    & 1.47 & $5.6\times10^{4}$ & $6.8\times10^{5}$ & $2.8\times10^{-14}$ & 1000 &7100 & $1.8\times10^{-9}$ & $7.3\times10^{-5}$\\ \hline
  \end{tabular}
  }
  \\
 \begin{flushleft}
   Note.\textemdash The parameters from left to right correspond to the core radius, core mass, 
   total mass, core angular velocity, far-UV intensity, temperature, 
   and chemical fractions of H$_{2}$, H$^{+}$, and e. 
 \end{flushleft}
 \end{center}
\end{table*}

The properties of the two clouds are summarized in Table~\ref{Tab:1}. 
The two clouds have almost the same mass, but the angular velocity of the core is larger in the {\it filamentary} cloud. 
For each cloud, we set our initial conditions using the spherically-averaged data of the three-dimensional simulations in \cite{Chon:2016}:
we set the initial surface density as 
\begin{align}
\Sigma(r) = {\displaystyle \int^{(r_{\mathrm{out}}^2-r^2)^{1/2}}_{-(r_{\mathrm{out}}^2-r^2)^{1/2}} \tilde{\rho} \left(\sqrt{r^2+z^2}\right)~\mathrm{d}z}~,
\label{Eq:30}
\end{align}
and the initial angular velocity as
\begin{align}
\Omega(r) = \frac{\tilde{v}_\phi(r)}{r}~,
\label{Eq:31}
\end{align}
where $\tilde{\rho}$ is the spherically averaged density as a function of $(r^2+z^2)^{1/2}$ 
and $\tilde{v}_\phi$ is the density-weighted spherical-average of the rotational velocity, 
which is almost identical to the rotational velocity in the equatorial plane 
because the density is larger in the equatorial plane than in the polar direction. 
Approximately, the density and the angular velocity are constant in the core, but decrease in proportion to $r^{-2}$ and $r^{-1/2}$ in the envelope, respectively.

The initial temperature and chemical fractions of H, H$_{2}$, H$^{+}$, and e are set to the values
obtained from the one-zone calculation when the number density reaches 10$^{5}$ cm$^{-3}$.
In our thermal and chemical models, 
we consider the effects of external radiation from a nearby galaxy. 
Following \cite{Chon:2016}, we set the values of the far-UV intensity $J_{21}$ to
5000 for the {\it filamentary} cloud and 1000 for the {\it spherical} cloud (see Equation~\ref{Eq:23}).

\section{Result}
\label{Sec:3}

Here, we show our simulation results for the {\it filamentary} and {\it spherical} clouds.
We present the time evolution of the disks around the central stars in Section~\ref{Sec:3-1}
and the growth of central stars due to the intermittent accretion from the disks in Section~\ref{Sec:3-2}. 
In Section~\ref{Sec:3-3}, we compare our results with a run using the barotropic temperature-density relation.

\subsection{Time evolution of the gravitationally unstable disk}
\label{Sec:3-1}

\begin{figure*}
 \begin{center}
 \begin{tabular}{c} 
  {\includegraphics[width=1.95\columnwidth]{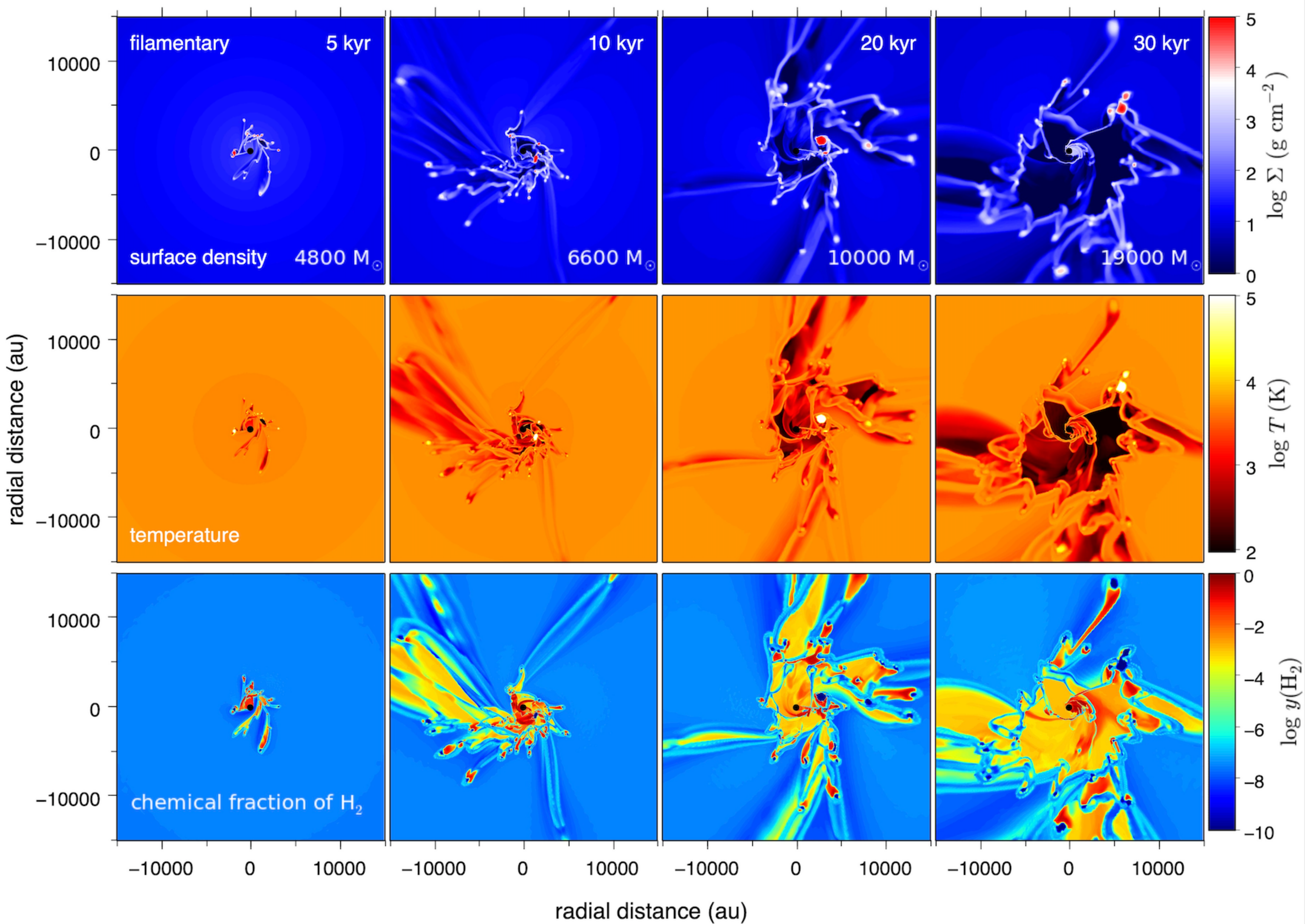}}
 \end{tabular}
 \caption{The time evolution of the disk in the {\it filamentary} cloud. 
 Each row corresponds to the surface density (top), temperature (middle), and chemical fraction of H$_2$ (bottom) at
 four different times, 5, 10, 20, and 30 kyr after the disk formation. 
 The central stellar mass at each time is shown in the bottom right corner of the upper panels.}
 \label{Fig:1}
 \end{center}
\end{figure*}
\begin{figure*}
 \begin{center}
 \begin{tabular}{c} 
  {\includegraphics[width=1.95\columnwidth]{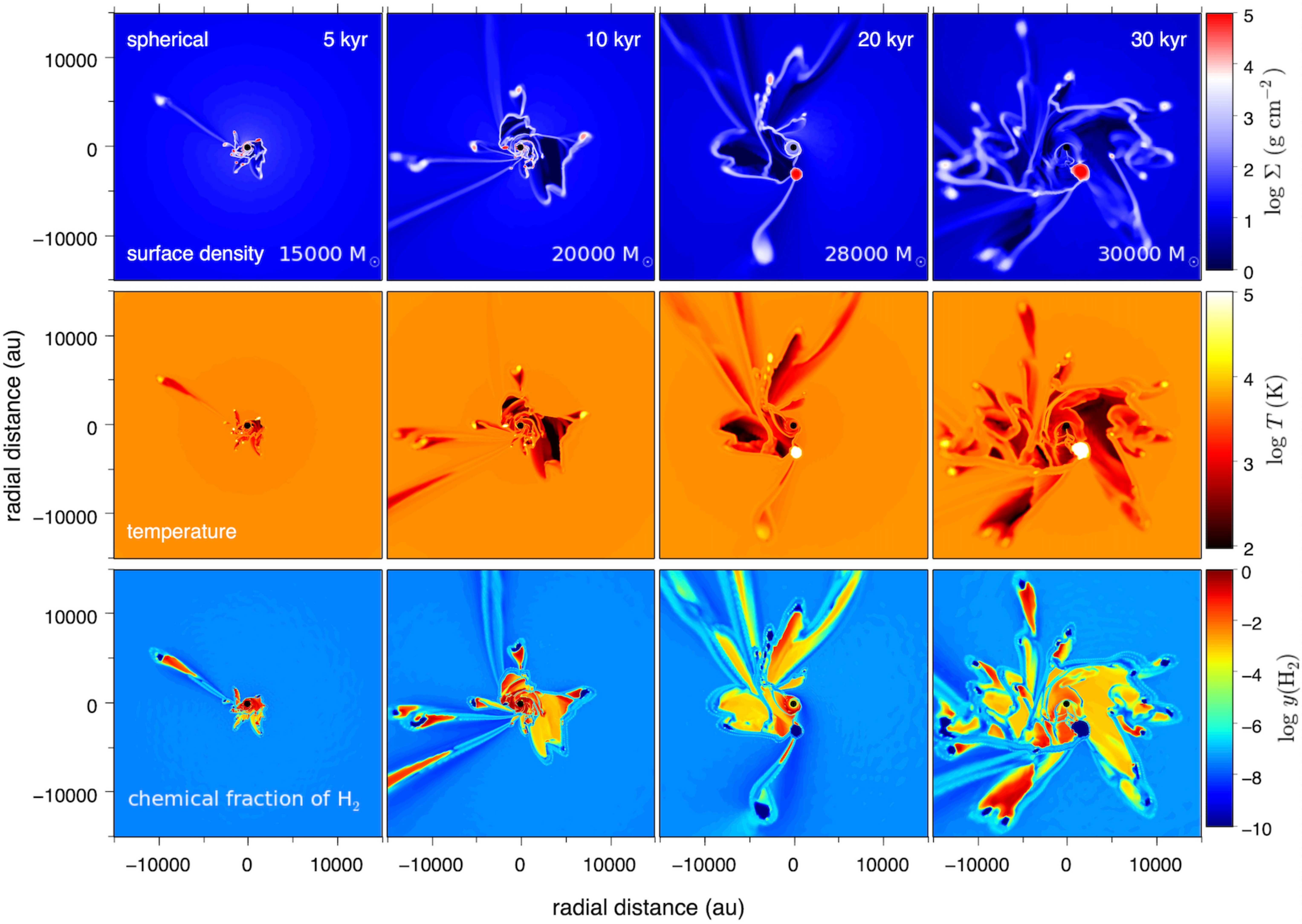}}
 \end{tabular}
 \caption{Same as Figure \ref{Fig:1}, but for the {\it spherical} cloud.}
 \label{Fig:2}
 \end{center}
\end{figure*}

Figure~\ref{Fig:1} shows the time evolution of the disk in the {\it filamentary} cloud.
In the figure, we present the surface density (top), the temperature (middle), and the chemical fraction of H$_2$ (bottom) 
at four different times, 5, 10, 20, and 30 kyr after the disk formation.

In the initial stage of gravitational collapse, 
the inner gas falls directly to the central sink cell 
because the angular momentum is lower at smaller radius in the initial condition.
As time passes, 
the infalling gas starts rotating around the sink and forming a disk
because the outer gas with high angular momentum hits the centrifugal barrier near the sink cell and cannot fall directly to the sink. 
The disk becomes massive and gravitationally unstable soon after its formation due to the large mass supply rate to the disk.
By 5 kyr after the disk formation, the gravitational instability leads to the formation of spiral arms and dozens of clumps, 
as seen in the surface density panel of Figure~\ref{Fig:1}. 
Most of the clumps are confined to a compact central area of 5000 au in the early phase (5 and 10 kyr),
but they later spread to a wider area of 10000 au, creating a central cavity region of 5000-10000 au  (20 and 30 kyr).
The clumps tend to rotate at outer radius as the angular momentum is brought in
by the gas supplied from the envelope.
The clumps form in the high-density parts of the spiral arms created due to the collisions of spiral arms.
Most clumps end up with falling down to the center,
maintaining the high accretion rate to the sink cell, as we will see in Section~\ref{Sec:3-2}.

The temperature of the envelope is quasi-isothermal with 5000-8000 K, consistent with
the one-zone calculation of a gravitationally collapsing core (Figure~\ref{Fig:B1}),
whereas that of the disk varies by three orders of magnitude (10$^2$-10$^5$ K) and largely different from the results of the one-zone calculation.
The temperature is closely related to the density structures:
it is high in the clumps (>10$^4$ K), 
low behind the spiral arms ($\sim$1000 K; the rotation is counterclockwise on the paper),
and even lower in the cavity region ($\sim10^{2}$ K; see the panels at 20 and 30 kyr).
The chemical fraction of H$_2$ is inversely correlated with the temperature: 
$y(\mathrm{H}_2)$ is $\sim10^{-6}$ in the envelope and spiral arms where the temperature is moderate,
smaller ($<10^{-10}$) in the hot clumps, 
and higher ($\gtrsim10^{-3}$) in the region behind the spiral arms and the cavity region where the temperature is low.

Figure~\ref{Fig:2} shows the time evolution of the disk in the {\it spherical} cloud,
which is qualitatively the same as in the {\it filamentary} cloud.
The disk is gravitationally unstable and fragmented to spiral arms and clumps,
whose distributions spread spatially with time.

In both runs, some clumps are ejected from the vicinity of the central star as a result of the gravitational interactions with the central star or other clumps. 
An ejected clump can survive as a single star if its velocity is larger than the escape velocity, 
\begin{align}
u_{\mathrm{esc}} &= \left( \frac{2GM_{\ast}}{R} \right)^{1/2}~\notag \\
&\simeq 23~\mathrm{km~s^{-1}}~\left( \frac{M_{\ast}}{3\times10^{4}~\mathrm{M}_{\odot}} \right)^{1/2}~\left( \frac{10^{5}~\mathrm{au}}{R} \right)^{1/2}~,
\label{Eq:32}
\end{align}
where $R$ is the radial distance of the ejected clump from the central star. 
For the {\it filamentary} cloud, the two clumps 
locating at $R=6$ and $8 \times 10^4$ au at the end of the calculation
have the velocities (35 and 47~km~s$^{-1}$, respectively) exceeding the escape velocity (Equation~\ref{Eq:32}), 
and would escape from the system thereafter. 
For the {\it spherical} cloud, on the other hand, no clump is found to have high enough velocity to escape. 

In the following, we give detailed analyses of the disk structures to
get a deeper understanding of the disk evolution.
Here, we present the analyses only for the {\it filamentary} cloud,
because those for the {\it spherical} cloud are similar, as expected from
the similar time evolution seen in Figures~\ref{Fig:1} and \ref{Fig:2}.

\begin{figure}
 \begin{center}
 \begin{tabular}{c} 
  {\includegraphics[width=0.95\columnwidth]{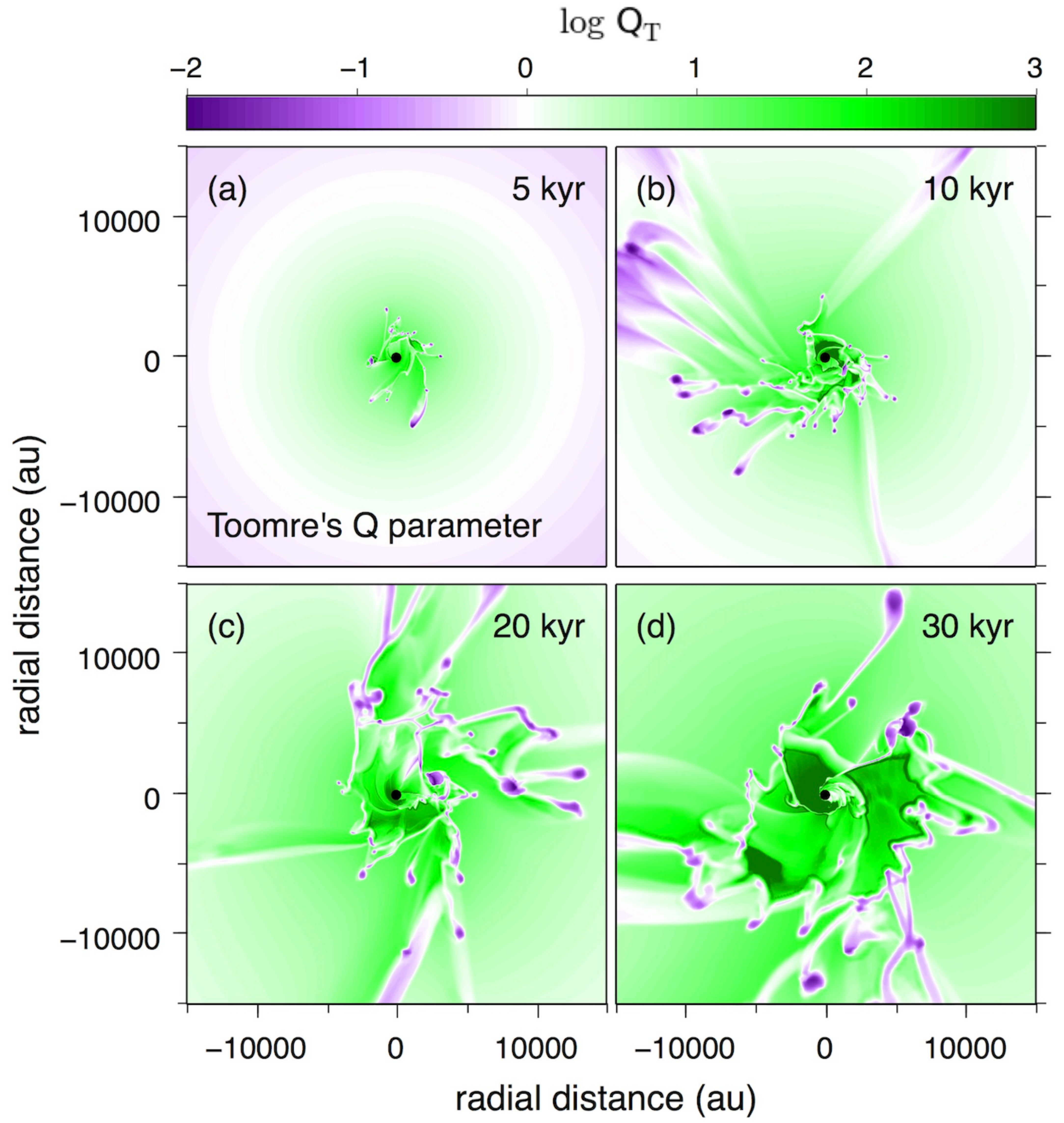}}
 \end{tabular}
 \caption{Spatial distributions of Toomre's Q parameter in the {\it filamentary} cloud.
  The time of each panel is the same as in Figure~\ref{Fig:1}.}
 \label{Fig:3}
 \end{center}
\end{figure}

In order to examine the gravitational instability of the disk, 
we plot in Figure~\ref{Fig:3} the spatial distributions of Toomre's Q parameter \citep{Toomre:1964}
\begin{align}
Q_{\mathrm{T}} = \frac{c_{\mathrm{s}}\Omega}{\pi G\Sigma}~, 
\label{Eq:33}
\end{align}
where we have replaced the epicyclic frequency with $\Omega=v_{\phi}/r$ assuming quasi-Keplerian rotation.
From the Toomre's criterion, the disk is gravitationally unstable in the region with $Q_{\mathrm{T}}<1$ (purple),
marginally stable in the region with $Q_{\mathrm{T}}=1$ (white), 
and stable in the region with $Q_{\mathrm{T}}>1$ (green).
It is clear from the comparison with Figure~\ref{Fig:1} that 
the distribution of $Q_{\mathrm{T}}$ is closely related to the distributions of the surface density and temperature (they are also closely related each other
as mentioned above):
the high-density regions (i.e., clumps) have $Q_{\mathrm{T}}<1$,
while the low-density regions has $Q_{\mathrm{T}}>1$;
the spiral arms have $Q_{\mathrm{T}}\approx1$, which means they are in the critical state of disk fragmentation. 
This value of $Q_{\mathrm{T}}$ along the spiral arms
confirms that the gravitational instability of disk in fact causes their formation.
The distribution of $Q_{\mathrm{T}}$ is consistent with a picture of
gravitationally unstable disks in which the gravitational torque of spiral arms and clumps drives
the accretion flows \citep[e.g.,][]{Matsukoba:2019}.

\begin{figure}
 \begin{center}
 \begin{tabular}{c} 
  {\includegraphics[width=0.95\columnwidth]{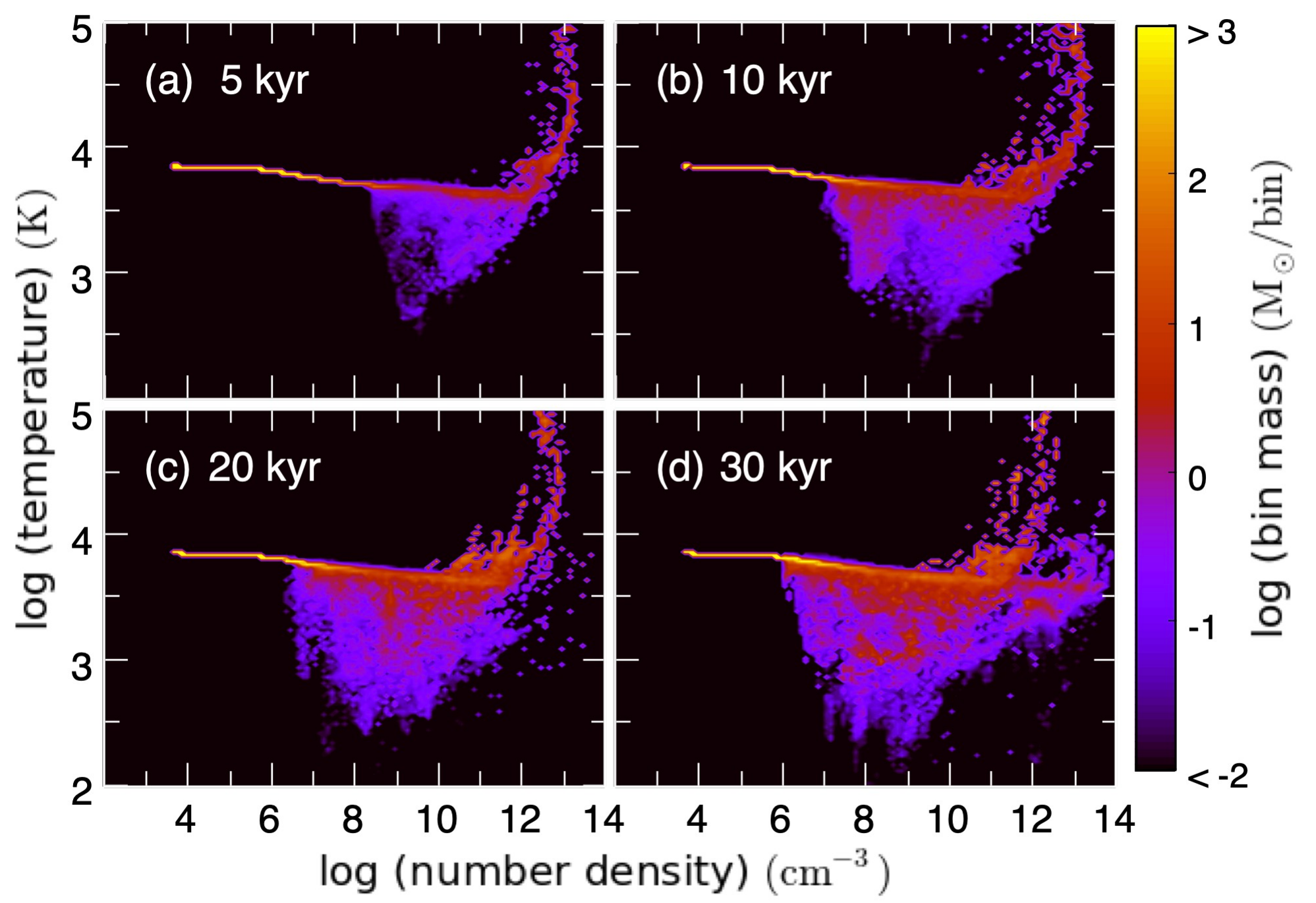}}
 \end{tabular}
 \caption{Gas mass distributions on the density-temperature phase diagrams for the {\it filamentary} cloud. 
 The color indicates the mass in each density-temperature bin with
 the widths of $\Delta~\mathrm{log}\,n_{\mathrm{H}}=0.1$ and $\Delta~\mathrm{log}\,T=0.025$.
  The time of each panel is the same as in the Figure~\ref{Fig:1}.}
 \label{Fig:4}
 \end{center}
\end{figure}

The gravitational instability of the disk depends on the temperature of the gas since $Q_{\mathrm{T}}\propto c_{\mathrm{s}}$ from Equation~\eqref{Eq:33}. 
To understand the thermal evolution of gas, 
we plot the mass distributions on the density-temperature phase diagrams in Figure~\ref{Fig:4}. 
In all four panels, a large amount of gas is distributed isothermally 
with the temperature $\sim5000-8000$ K between the number density $\sim10^{3}$  cm$^{-3}$ and $10^{8}$ cm$^{-3}$.
This is the envelope contracting due to the atomic hydrogen cooling. 
The low-temperature (< 1000 K) regions
with the number density 10$^{7}$-10$^{11}$ cm$ ^{-3}$ 
correspond to the regions behind the spiral arms. 
When the spiral arms pass through and sweep out the gas,
not only the density but also the temperature significantly
decreases. The decrease of the temperature is roughly adiabatic  ($T\propto \rho^{2/3}$) because
the expansion cooling works as the main coolant.
Other features evident in the figure are the high density ($> 10^{11}~\mathrm{cm}^{-3}$) and high temperature ($>10^{4}$ K) regions 
that correspond to the optically-thick clumps heated due to adiabatic contraction.

\begin{figure}
 \begin{center}
 \begin{tabular}{c} 
  {\includegraphics[width=0.9\columnwidth]{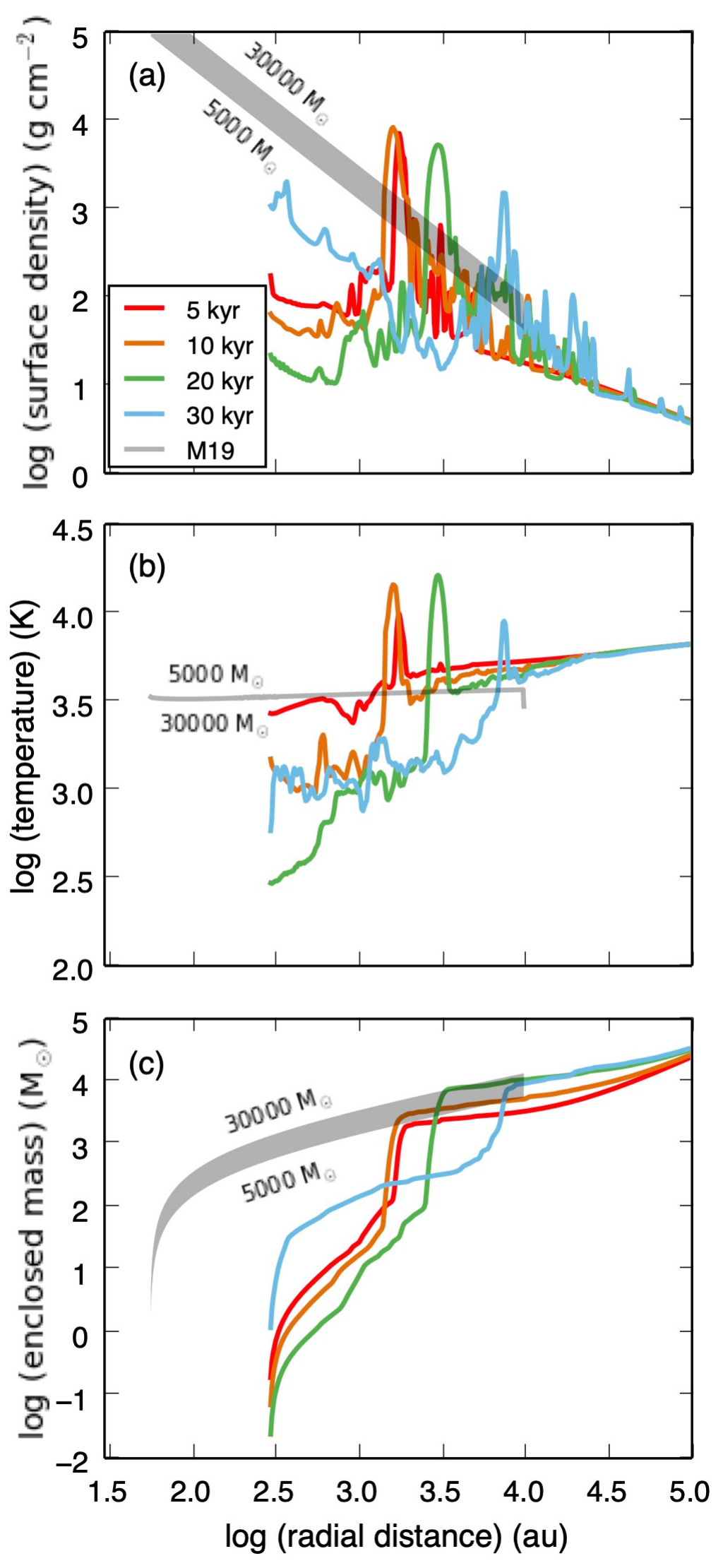}}
 \end{tabular}
 \caption{Radial profiles of the azimuthally-averaged (a) surface density and (b) temperature and (c) the enclosed mass in the {\it filamentary} cloud. 
 The gray filled lines show the radial profiles of the one-dimensional steady accretion disk model \citep{Matsukoba:2019} for the stellar mass between 5000 M$_{\odot}$ and 30000 M$_{\odot}$. 
 The colors indicate the times after the disk formation, 5 kyr (red), 10 kyr (orange), 20 kyr (green), and 30 kyr (blue),
 when the stellar masses are 4800~M$_{\odot}$, 6600~M$_{\odot}$, 10000~M$_{\odot}$, and 19000~M$_{\odot}$, respectively. }
 \label{Fig:5}
 \end{center}
\end{figure}

Next, we examine the one-dimensional structure of the disk. 
The radial profiles of the azimuthally-averaged (a) surface density and (b) temperature and (c) the enclosed mass in the {\it filamentary} cloud 
are shown in Figure~\ref{Fig:5}. 
Along with the profiles at 5 (red), 10 (orange), 20 (green), and 30 (blue) kyr after the disk formation, 
we plot the radial profiles of the one-dimensional steady accretion disk model in \cite{Matsukoba:2019} with the gray filled lines,
for which we set the two parameters of the model, the stellar mass and accretion rate, 
to 5000-30000~M$_{\odot}$ and 0.1~M$_{\odot}$~yr$^{-1}$, respectively.
In this one-dimensional model, we solve the non-equilibrium chemical and thermal evolution assuming
that the disk is marginally unstable with $Q_{\mathrm{T}}=1$.
Here, we adjust the the outer edge of the disk to 10$^{4}$ au (it was 10$^{3}$ au in \citealt{Matsukoba:2019}), 
but otherwise adopt the same set-up as in \cite{Matsukoba:2019}.

At each time, we see a strong density peak with $10^{3}-10^{4}$~g~cm$^{-2}$ at $10^{3}-10^{4}$ au (Figure~\ref{Fig:5}a),
which is coincided with a temperature peak with $\gtrsim10^{4}$~K (Figure~\ref{Fig:5}b).
The peak corresponds to the largest clump at each time, which is seen as the largest red clump in
each panel of the surface density snapshots in Figure~\ref{Fig:1}.
These clumps are actually the identical clump observed at a different time, 
which we have confirmed from the snapshots with short time intervals.
The clump mass, which can be estimated from the jumps in the enclosed mass profile (Figure~\ref{Fig:5}c),
is $\sim$1000~M$_{\odot}$ at 5 kyr and grows to $\sim$10000~M$_{\odot}$ at 30 kyr,
as a result of mergers with other clumps and accretion of surrounding gas.
While the clump grows in mass, it also acquires the angular momentum through the growth process,
and thus its separation from the center gradually expands, as indicated by
the position of the peak moving outward with time in Figure~\ref{Fig:5}.
Similar orbital evolution was reported in the simulations of Pop III star formation \citep{Chon:2019, Sugimura:2020}.

Now, let us briefly compare the simulation results with the one-dimensional steady accretion disk model.
In Figure~\ref{Fig:5} (a and b), 
the radial profiles of surface density and temperature are roughly consistent with
the one-dimensional model outside the peaks, 
but largely different inside the peaks, 
where the surface density is 2-3 orders of magnitude smaller than that of the one-dimensional model and 
the temperature drops from $\sim$3000 K to 1000 K due to the expansion cooling. 
This lower surface density implies that 
the gap opening is induced by the gravitational interaction of the central star, the largest clump, and infalling gas.
The one-dimensional model fails to reproduce the simulation results
because such effect is not taken into account. 

\begin{figure}
 \begin{center}
 \begin{tabular}{c} 
  {\includegraphics[width=0.95\columnwidth]{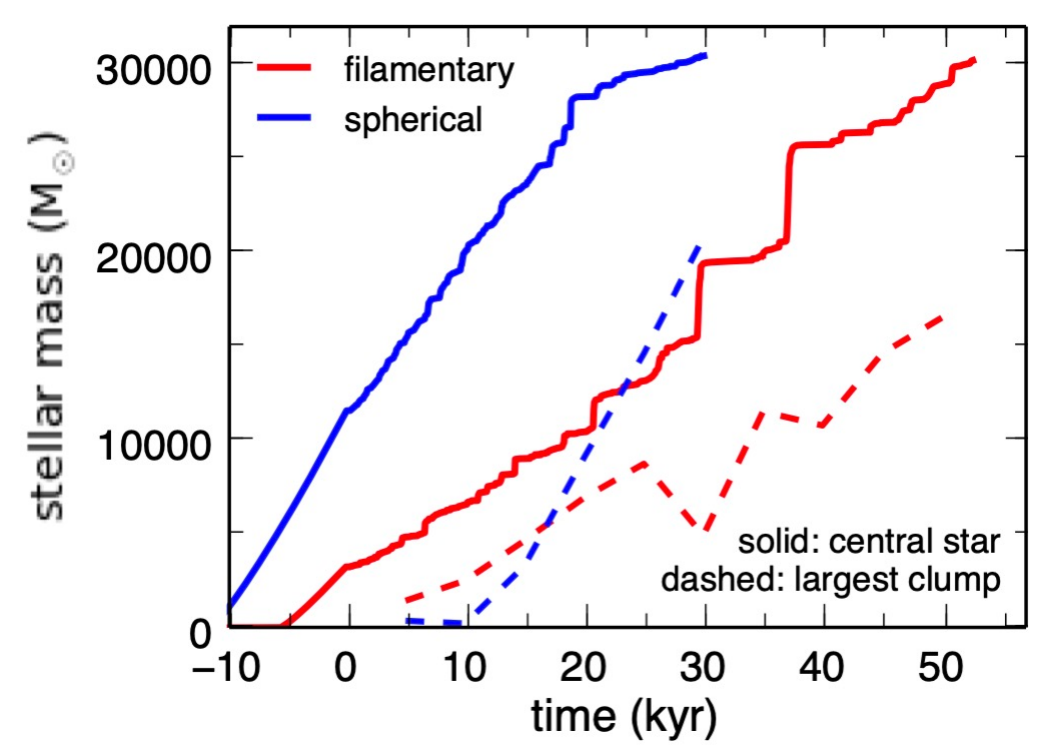}}
 \end{tabular}
 \caption{Time evolution of the star and the largest clump masses. The colors correspond to {\it filamentary} cloud (red) and {\it spherical} cloud (blue). 
 The solid and dashed lines represent the central stellar mass and the largest clump mass, respectively.}
 \label{Fig:6}
 \end{center}
\end{figure}

Before closing this section, it is worth noticing that massive clumps 
are formed in both the {\it filamentary} and {\it spherical} clouds (see the upper-right panels in Figures~\ref{Fig:1} and \ref{Fig:2}). 
We show the mass evolution of the central star and the largest clump in Figure~\ref{Fig:6}. 
Here the mass of the largest clump is calculated by summing the mass in the grids with the surface density above 10$^4$ g cm$^{-2}$ around the maximum density in the clump, which is sampled at every 5 kyr starting from 5 kyr after the disk formation. 
The largest clump has grown to 17000 ({\it filamentary}) and 21000 ({\it spherical})~M$_{\odot}$ 
and locates at $\sim10^4$ ({\it filamentary}) and $~3\times10^3$ ({\it spherical}) au away from the central
star at the end of the calculations (at $\sim$50 and 30 kyr after the disk formation, respectively), 
when the central stellar masses reach 30000~M$_{\odot}$. 
The largest clump in each run potentially makes a binary stellar system with the central star eventually (see also Section~\ref{Sec:4}).
Although longer timescale calculation is required to draw definite conclusion, previous simulations
in a similar context have observed 
the formation of binary SMSs \citep{Chon:2018,Latif:2020}.

\subsection{Stellar evolution under intermittent accretion}
\label{Sec:3-2}

\begin{figure}
 \begin{center}
 \begin{tabular}{c} 
  {\includegraphics[width=0.95\columnwidth]{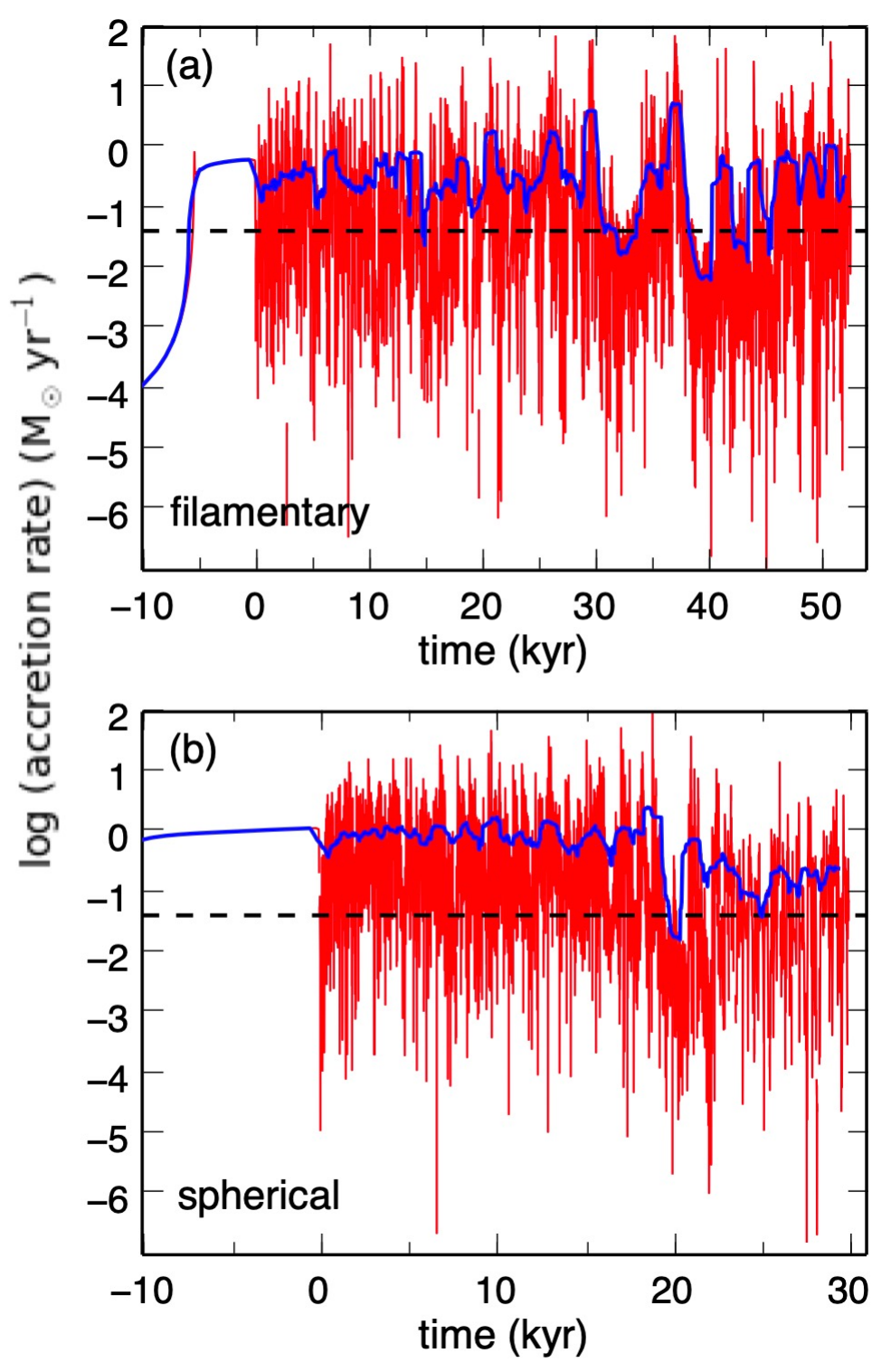}}
 \end{tabular}
 \caption{Accretion histories onto the central star with different initial conditions, (a) {\it filamentary} cloud and (b) {\it spherical} cloud. 
 The red line represents the raw accretion rate, and the blue line denotes the time-averaged rate with bin of 1000 years. 
 The black dashed line indicates the critical rate ($4\times10^{-2}~\mathrm{M}_{\odot}~\mathrm{yr}^{-1}$),
  below which the star begins to emit ionizing photons due to stellar contraction if the accretion rate is constant.}
 \label{Fig:7}
 \end{center}
\end{figure}

In Figure~\ref{Fig:6}. 
The stellar masses reach the final mass of 30000~M$_{\odot}$ in both cases, but 
the growth time is shorter in the {\it spherical} cloud than in the {\it filamentary} cloud ($\sim$30 kyr and $\sim$50 kyr, respectively).
The central star accretes the gas more rapidly in the {\it spherical} cloud,  because
the {\it spherical} cloud has the smaller initial angular momentum than the {\it filamentary} cloud.

Figure~\ref{Fig:7} shows the accretion rates in the two cases.
We plot the time-averaged rates with bins of 1000 years (blue) along with
the raw rates (red).
The raw accretion rates violently fluctuate by nine orders of magnitude in both cases,
while the averaged rates fluctuate much more gently with some occasional strong bursts.
We attribute the strong fluctuations of the averaged rates to the interaction with the massive clumps
that have the masses comparable to the central stars, as explained in Section~\ref{Sec:3-1}.
In contrast, the fluctuation of the averaged rate is especially small in the early time ($\lesssim$15 kyr)
in the {\it spherical} cloud, partly because clumps as massive as the central star has yet to form for this period.
The massive clumps exert gravitational torque on the gas in the disks, causing accretion bursts that are followed by short quiescent periods. 
Besides, they sometimes approach the central star so closely as to be tidally disrupted and some of their material is transferred to 
the central star, as we observe in the snapshots with short time intervals. Such events cause
the particularly large accretion bursts at $\sim$30 and 38 kyr in the {\it filamentary} cloud,
which increase the stellar mass by $\sim$5000~M$_{\odot}$ (see also Figure~\ref{Fig:6}). 

As described in the introduction, the radiative feedback by the ionizing radiation from the protostars
may quench the accretion if the accretion rate drops below the critical value of
4$\times10^{-2}~\mathrm{M}_{\odot}~\mathrm{yr}^{-1}$ \citep[black dashed line in Figure~\ref{Fig:7};][]{Hosokawa:2012, Hosokawa:2013} and cannot keep the stellar surfaces inflated.
According to \cite{Sakurai:2015},
if a quiescent period $\Delta t_{\mathrm{q}}$, for which the accretion rate is below the critical value,
is longer than the KH timescale at the stellar surface $t_{\mathrm{KH,surf}}$ (Equation~\ref{Eq:1}),
the radiative feedback quenches the accretion
because the stellar surface shrinks significantly and the associated rise
of the effective temperature
leads to the emission of strong ionizing radiation.
Conversely, if $\Delta t_{\mathrm{q}}<t_{\mathrm{KH,surf}}$, 
the radiative feedback is ineffective because
the contracting stellar surface turns to inflating
due to the revival of the accretion rate before the strong ionizing radiation is emitted.

Below we estimate the effect of radiative feedback in our simulated cases, using the above condition. 
In the {\it filamentary} cloud, 
the longest quiescent period $\Delta t_{\mathrm{q}}\sim2000$ yr at $t=$38 kyr
is shorter than the KH timescale $t_{\mathrm{KH,surf}}\sim7000$ yr for the stellar mass of $\sim$25000~M$_{\odot}$ at this time (Equation~\ref{Eq:1}). 
Similarly, in the {\it spherical} cloud, 
the longest quiescent period $\Delta t_{\mathrm{q}}\sim800$ yr at $t=$20 kyr
is shorter than the KH timescale $t_{\mathrm{KH,surf}}\sim7500$ yr for the stellar mass of $\sim$28000~M$_{\odot}$ at this time.
There are other quiescent periods with $\Delta t_{\mathrm{q}}\sim700$ yr at $t=33,~44,~$and $46$ kyr in the {\it filamentary} cloud,  but they are all about one order of magnitude shorter than the KH timescales.
Therefore, in our simulated cases, the quiescent periods never exceed the KH timescales, 
and thus we conclude that the radiative feedback by ionizing radiation,
although not explicitly considered in our simulations, does not affect the accretion flows.

In the both runs studied, the largest clump is as massive as the central star. 
The gas surrounding that clump, however, is not affected so much by the radiative feedback from the star 
formed there because the accretion rate is higher than the critical value: 
0.1 for the {\it filamentary} and 0.2 M$_{\odot}~\mathrm{yr}^{-1}$ for the {\it spherical} cloud
from Figure~\ref{Fig:6}.

\subsection{Comparison with the calculation using a barotropic relation}
\label{Sec:3-3}

\begin{figure*}
 \begin{center}
 \begin{tabular}{c} 
  {\includegraphics[width=1.95\columnwidth]{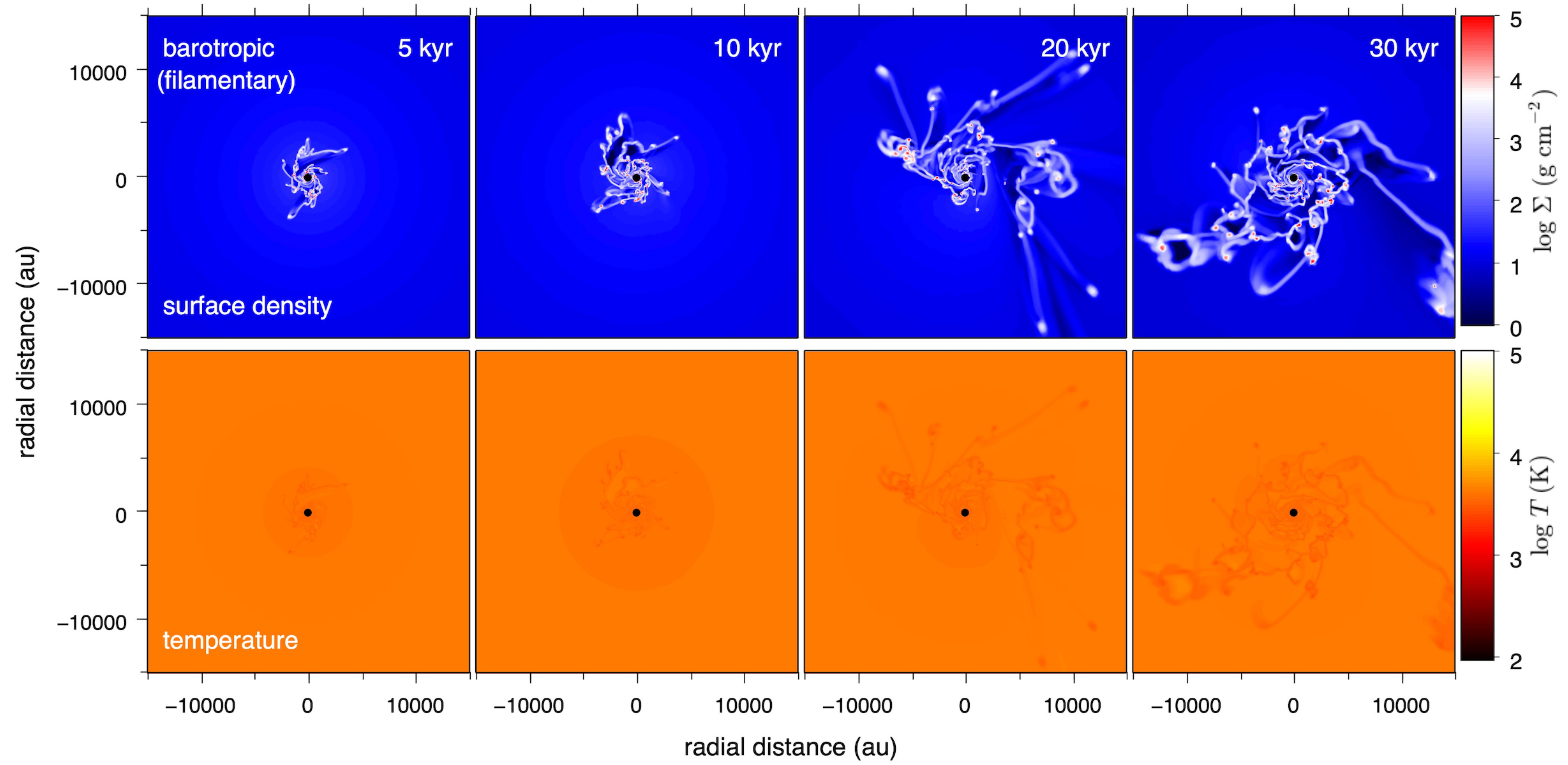}}
 \end{tabular}
 \caption{Same as Figure~\ref{Fig:1}, but for the run with a barotropic relation
 starting from the initial condition of the {\it filamentary} cloud.
The spatial distributions of surface density (upper) and temperature (lower) are shown.}
 \label{Fig:8}
 \end{center}
\end{figure*}

In this section, we compare our main run described above
with an additional run using a barotropic temperature-density relation,
as in the previous study \citep{Sakurai:2016}, focusing on the
case of the {\it filamentary} cloud.
Figure~\ref{Fig:8} shows the spatial distributions of the surface density and temperature in the run 
starting from the same initial condition of the {\it filamentary} cloud
but using the barotropic relation instead of solving the energy equation.

In the upper panels of Figure~\ref{Fig:8},
the circumstellar disk is fragmented into a large number of spiral arms and clumps from an early stage 
and their number further increases with time.
While the runs with our thermal model and the barotropic relation
commonly show the fragmentation of the disks into spiral arms and clumps,
we find two major differences regarding the properties of the clumps:
(1) the number of clumps in the run with the barotropic relation is larger than in the run with our thermal model,
and (2) the massive clumps found in the run with our thermal model is not found in the run with the barotropic relation.
The dependence of the characteristics of clumps on the adopted thermal model was 
also argued in the case of Pop III star formation \citep{Clark:2011}.

We attribute these differences mainly to the lack of resolution in the run with the barotropic relation. 
The local Jeans length must be resolved by at least four grids in order to avoid artificial fragmentation \citep{Truelove:1997}. 
In the barotropic run, however, we find that this condition is not satisfied near clumps. 
With the barotropic relation, the local Jeans length 
at (10$^{16}$~cm$^{-3}$, 7000K),  
where the gas becomes adiabatic, is
\begin{align}
\lambda_{\mathrm{J}} = 1.3~\mathrm{au}~\left(\frac{n_{\mathrm{H}}}{10^{16}~\mathrm{cm}^{-3}}\right)^{-1/2}~\left(\frac{T}{7000~\mathrm{K}}\right)^{1/2}.
\label{Eq:34}
\end{align}
In late stages of the run, 
clumps are distributed within around 5000 au from the central star, where
the grid size is 80 au. This means the local Jeans length around the clump location is far below the resolution: the required number of grids for the \cite{Truelove:1997} criterion is more than 250 times that in our calculations. 
Consequently, artificial fragmentation of the clumps sometimes takes place. 

The temperature distributions in the run with the barotropic relation, as shown in the lower panels of Figure~\ref{Fig:8}, 
is largely different from those in the run with our thermal model (Figure~\ref{Fig:1}).
In the run with our thermal model, the temperature varies by three orders of magnitude (100-10$^5$ K), 
mainly due to the compressional/shock heating and the expansion cooling associated with the dynamics of spiral arms and clumps.
In the run with the barotropic relation, however, the gas remains almost isothermal with $\sim$5000-8000 K 
because the barotropic relation is calculated without taking into account the thermal processes associated with the gas dynamics in the disk.

\begin{figure}
 \begin{center}
 \begin{tabular}{c} 
  {\includegraphics[width=0.95\columnwidth]{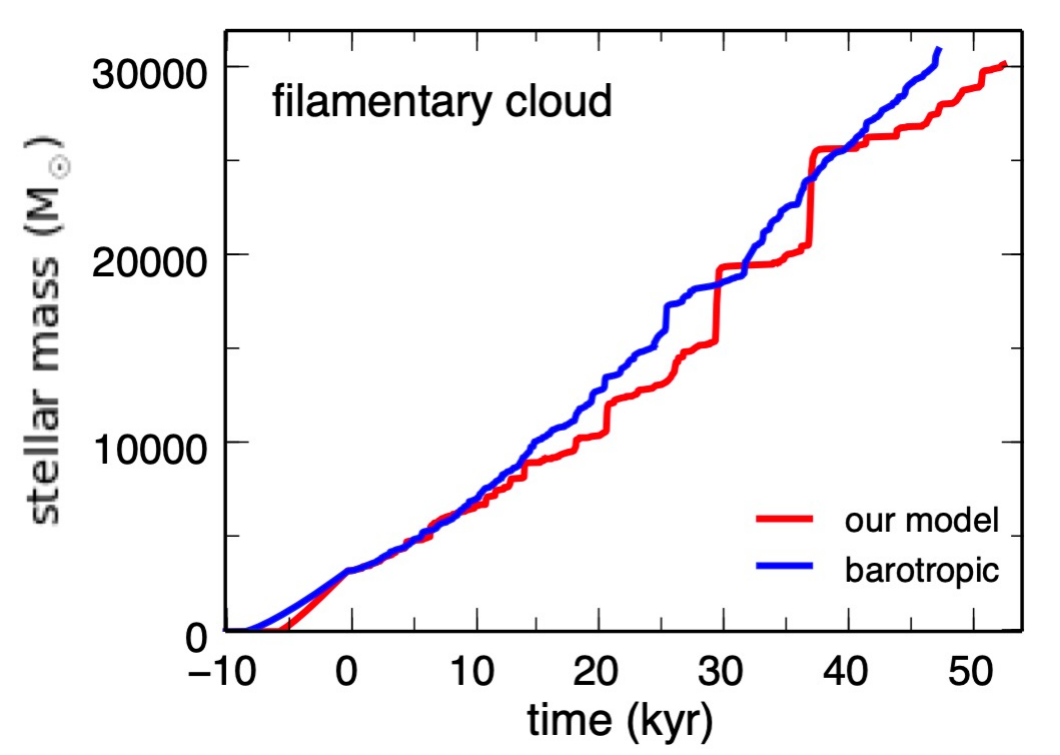}}
 \end{tabular}
 \caption{The dependence of time evolution of stellar mass on the treatment of thermal evolution.
  We show the results from the runs starting from the initial condition of the {\it filamentary} cloud
with our thermal model (red) and the barotropic relation (blue).
}
 \label{Fig:9}
 \end{center}
\end{figure}
\begin{figure}
 \begin{center}
 \begin{tabular}{c} 
  {\includegraphics[width=0.95\columnwidth]{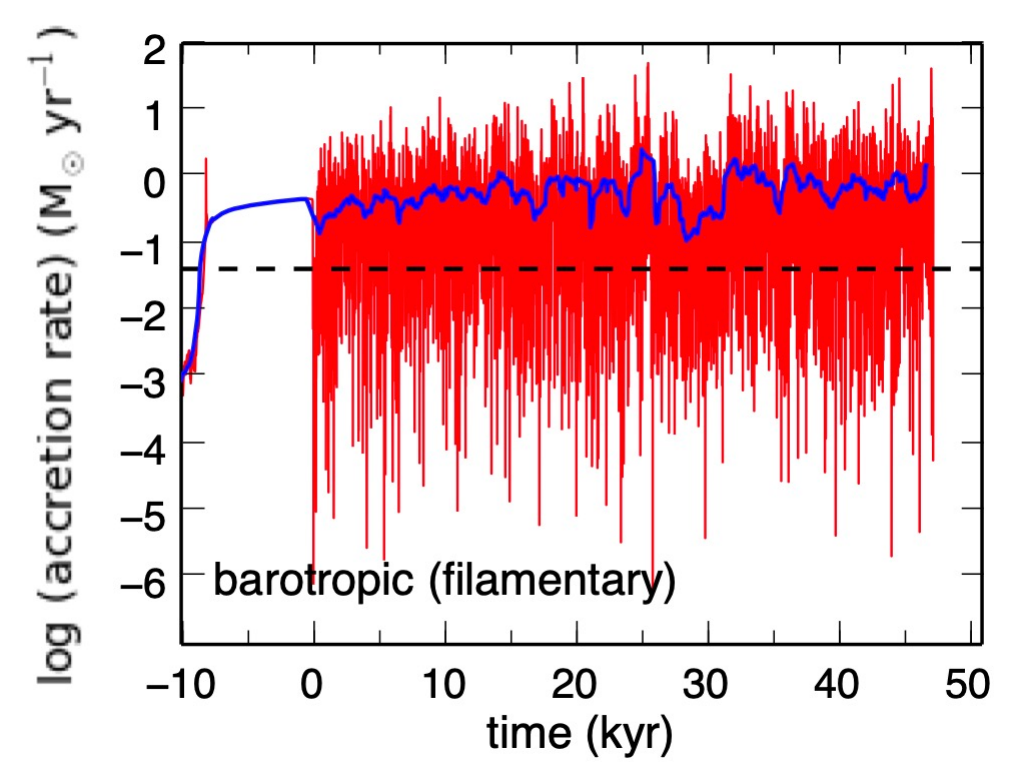}}
 \end{tabular}
 \caption{Same as Figure~\ref{Fig:7}, but for the run starting from the initial condition of the {\it filamentary} cloud with the barotropic relation.}
 \label{Fig:10}
 \end{center}
\end{figure}

In Figure~\ref{Fig:9}, we compare the time evolution of the central stellar masses 
in the runs  with our thermal model (red) and the barotropic relation (blue).
We also provide the time evolution of the accretion rate in the run with the barotropic relation
in Figure~\ref{Fig:10} 
(see Figure~\ref{Fig:7}a for the run with our thermal model). 
While the stellar masses reach 30000 M$_{\odot}$ around the same time ($\sim$50 kyr) in both runs,
the mass growth is smoother and the time-averaged accretion rate never falls 
below the critical rate in the run with the barotropic relation, because 
smaller but more numerous clumps are formed and continuously accrete onto the central star. 
This implies that runs with the barotropic relation 
underestimate the length of quiescent periods.
Although the quiescent periods are shorter than the KH timescales in our examined cases,
as explained in Section~\ref{Sec:3-2},
the radiative feedback still potentially prevents the accretion in some cases.
In such cases, the use of the barotropic relation may lead to a wrong conclusion on
the role of the radiative feedback. 
Therefore, realistic treatment of the thermal evolution 
is crucial to understand the SMS formation.

\section{Summary and Discussion}
\label{Sec:4}

Supermassive stars (SMSs) are prominent candidate objects 
for the origin of supermassive black holes (SMBHs) observed in the early Universe. 
In this paper, we have investigated the time evolution of the disks around
growing SMSs by performing vertically-integrated two-dimensional 
hydrodynamic simulations starting from two cosmological initial conditions named {\it filamentary} and {\it spherical} clouds \citep{Chon:2016}.
We have put a particular focus on the time variation of the accretion rate,
because it was known 
that the ionizing radiation from a protostar can terminates the gas accretion, and hence the stellar growth,
if the quiescent period of the intermittent accretion is longer than the Kelvin-Helmholtz (KH) timescale at the stellar surface \citep{Sakurai:2015}.

In both the {\it filamentary} and {\it spherical} clouds, gravitationally unstable circumstellar disks that are
fragmented into spiral arms and clumps 
provide the central stars with gas in an intermittent way.
The longest quiescent periods are 2000 ({\it filamentary}) and 800 ({\it spherical}) years 
and shorter than the KH timescales of 7000  ({\it filamentary}) and 7500 years  ({\it spherical}), respectively,
suggesting that protostars can continue to grow until they become SMSs without affected by the ionizing radiation. 
By the time the central star have grown to 30000~M$_{\odot}$, 
the largest clump around it reaches 17000~({\it filamentary}) and 21000~M$_{\odot}$ ({\it spherical}) , respectively.
The system may evolve to a binary SMS
and eventually become a binary BH.

Furthermore, we have compared our results with 
an additional run adopting the same initial condition 
but using the barotropic temperature-density relation,
as in the previous work \citep{Sakurai:2016}.
In this run, the quiescent periods are shorter because 
smaller but more numerous clumps are formed and continuously accrete onto the central star.
Thus, we have found that without solving the thermal and chemical evolution, one
tends to underestimate the length of quiescent periods and
may come to a wrong conclusion on the role of the radiative feedback.
Moreover, although we have observed the formation of binary SMSs
in both of the runs with our thermal model,
only small clumps form in the run with the barotropic relation.
From these reasons, we conclude that the simulations using the barotropic relation cannot 
describe the actual formation processes of SMSs. 

\cite{Chon:2018} studied the SMS formation using
three-dimensional simulations with the same initial conditions as ours.
They observed the formation of only 25 ({\it filamentary}) and 13 ({\it spherical}) clumps in each cloud, 
although we have observed the formation of more than hundred clumps in each cloud. 
Below, we provide three effects that probably play some roles in causing this difference.
Firstly, our simulations have higher effective resolution 
than their smoothed-particle hydrodynamic simulations. 
In \cite{Chon:2018}, they assumed that the gas becomes adiabatic at the density 
higher than 10$^{13}$ cm$^{-3}$ to save the computational costs,
effectively setting the minimum resolution of about 40 au.
As our minimum grid size is 5 au near the inner boundary at 300 au,
we can follow the formation of smaller clumps in the inner region.
Secondly, gravitational instability was suppressed in \cite{Chon:2018}
by the higher disk temperature than in our simulations.
Since they did not consider the H$^-$ free-bound emission, which is the primary cooling process in the disk, 
the disk temperature was higher than ours. 
Finally, dense parts of spiral arms that are supposed to fragment into clumps
are more easily formed in our simulations,
because in two-dimensional simulations, the vertically extended structures are confined to the disk plane
and the density increases associated with the collision of spiral arms may be overestimated.
Recently, \cite{Latif:2020} studied the long-term ($\sim$1 Myr) evolution of forming SMSs 
using three-dimensional adaptive mesh refinement simulations.
They also found the formation of multiple clumps, but their number is only ten or less in each run
partly because their resolution was much worse than ours with the minimum grid size of 2000 au. 
\cite{Regan:2020} also performed the three-dimensional simulations with similar resolution in \cite{Latif:2020} and 
found more than 20 massive stars with >1000~M$_{\odot}$. 
Unlike in our runs, however, 
those stars are formed via the fragmentation of the cloud core rather than via the disk fragmentation.

In each run, a clump reaches a comparable mass with the central star. 
Its orbital distance from the central star is 2000 au in the early phase ($\sim$ 5 kyr) 
and gradually increases with time, finally reaching 9000 ({\it filamentary}) and 4000 au ({\it spherical}), respectively. 
We expect the separation will increase even after that
owing to the acquisition of angular momentum by the gas accretion, as suggested in recent simulations of binary accretion \citep{Duffell:2020, Munoz:2020}.
In fact, long-term simulations in \cite{Latif:2020} demonstrated the formation of binary SMSs with a wide separation ($\sim$pc). 
The massive clumps in our runs may also make binary SMSs with their central stars. 
If such a binary system 
survives without merger until the end of 
the SMS lifetime, 
the outcome will be a binary BH system with
$\gtrsim10000~\mathrm{M}_{\odot}$ \citep{Umeda:2016}.
The merger of such binary BHs is particularly important because
the gravitational waves from the merger event will be detectable 
by next-generation gravitational wave detectors, 
e.g., {\it Deci-hertz Interferometer Gravitational wave Observatory} \citep[DECIGO:][]{Kawamura:2011} 
and {\it Laser Interferometer Space Antenna} \citep[LISA:][]{Amaro-Seoane:2012}. 
As the accretion onto each star and the associated orbital evolution of the binaries
still continue at the end of our simulations,
it is necessary to carry out long-time simulations to address the properties of the binary BHs.

Our numerical results depend somewhat on the resolution because we cut off the cooling 
at high-density regions using Equations~(\ref{Eq:28}) and~(\ref{Eq:29}). 
In order to examine the effect of the resolution, 
we have carried out the additional runs with 256$\times$256 and 768$\times$768 grids 
(while with 512$\times$512 grids in our runs so far) 
until 10 kyr after the disk formation. 
We found that the number of small fragments increases toward higher resolution, while 
a binary star system emerges at the center in the both runs. 
The quiescent period is at most $\sim$100 yr in both runs and always shorter than the KH timescale. 
The length of quiescent period does not change with the resolution 
because the number of small fragments does not change so much the quiescent period as we mentioned in Section~\ref{Sec:3-3}. 

Among the effects not considered in this work, 
the increase of stellar spin due to the accumulation of the angular momentum of accreted gas 
may play a role in ceasing the stellar growth.
To maintain the accretion, the sum of the radiative and centrifugal forces must be smaller than the gravity on the stellar surface,
which is known as the $\Omega\Gamma$ limit (see, e.g., \citealt{Maeder:2000}, \citealt{Lee:2016}, \citealt{Takahashi:2017}; \citealt{Haemmerle:2018-1}). 
We need to investigate the angular momentum transport at the interface of disks and stellar surfaces, 
to follow the stellar spin evolution and understand the role of the $\Omega\Gamma$ limit in the SMS formation.

Although our simulations have followed the formation process of SMSs for $\sim$30-50 kyr, 
longer-time ($\sim$Myr) simulations are needed to decide the fate of growing SMSs. 
Moreover, three-dimensional simulations are needed to consider vertical gas dynamics  missed in our simulations.
In future studies, we will come back to high resolution long-term  three-dimensional simulations,
to reveal the true nature of SMS formation.


\section*{Acknowledgments}
RM acknowledges financial support from the Graduate Program on Physics for Universe of Tohoku University. 
E. I. V. acknowledges support from the Austrian Science Fund (FWF) under research grant P31635-N27. 
KS appreciates the support by the Fellowship of the Japan Society for the Promotion of Science for Research Abroad. 
This work is financially supported
by the Grants-in-Aid for Basic Research by the Ministry of
Education, Science and Culture of Japan (SC:19J00324, TH:19H01934, KO:17H02869, 17H01102, 17H06360). 
The numerical simulations were carried out on XC50 at the Center for Computational Astrophysics (CfCA) of National Astronomical Observatory of Japan.


\section*{Data availability}
The data underlying this article will be shared on reasonable request to the corresponding author.



\appendix

\section{Chemical reactions}
\label{App:a}

We follow the compositional evolution of 5 species, H, H$_2$, H$^+$, H$^-$, and e, 
by solving the non-equilibrium kinetic equations. 
The 22 reactions included in our chemical network are summarized with their rate coefficients in Table~\ref{Tab:a1}. 
In each row where two reaction numbers are given, the first and second numbers correspond to the forward and reverse reactions, respectively.
To obtain the rate coefficients for the reverse reactions, we use the method described in Appendix C of \cite{Matsukoba:2019}.

\begin{table*}
 \begin{center}
 \caption{Chemical reactions}
 \label{Tab:a1}
  \scalebox{1.0}[1.0]{ 
  \begin{tabular}{c l l c} \hline \hline
    Number & Reaction & Rate coefficient of forward reaction ($\mathrm{cm}^{3} \  \mathrm{s^{-1}}$) & Reference \\ \hline
    1, 2 & $\mathrm{H} + \mathrm{e} \rightleftharpoons \mathrm{H}^{+} + 2 \mathrm{e}$ & $k_{1} = \mathrm{exp}[-3.271396786 \times 10^{1}$ & {\small \cite{Janev:1987}} \\ 
    & & $~~~~~~ + 1.35365560 \times 10^{1} \  \mathrm{ln}~T_{\mathrm{e}} - 5.73932875 \times 10^{0} \  (\mathrm{ln}~T_{\mathrm{e}})^{2}$ & \\
    & & $~~~~~~ + 1.56315498 \times 10^{0} \  (\mathrm{ln}~T_{\mathrm{e}})^{3} - 2.87705600 \times 10^{-1} \  (\mathrm{ln}~T_{\mathrm{e}})^{4}$ & \\
    & & $~~~~~~ + 3.48255977 \times 10^{-2} \ (\mathrm{ln}~T_{\mathrm{e}})^{5} - 2.63197617 \times 10^{-3} \  (\mathrm{ln}~T_{\mathrm{e}})^{6}$ & \\
    & & $~~~~~~ + 1.11954395 \times 10^{-4} \ (\mathrm{ln}~T_{\mathrm{e}})^{7} - 2.03914985 \times 10^{-6} \  (\mathrm{ln}~T_{\mathrm{e}})^{8}]$ & \\
    3, 4 & $\mathrm{H}^{-} + \mathrm{H} \rightleftharpoons \mathrm{H}_{2} + \mathrm{e}$ & 
       $k_{3} = 1.3500\times10^{-9}( T^{9.8493\times10^{-2}}+3.2852\times10^{-1}T^{5.5610\times10^{-1}}$ & {\small \cite{Kreckel:2010}} \\    
    & & $~~~~~~ +2.7710\times10^{-7}T^{2.1826})/(1.0+6.1910\times10^{-3}T^{1.0461}$ & \\
    & & $~~~~~~ +8.9712\times10^{-11}T^{3.0424}+3.2576\times10^{-14}T^{3.7741})$ & \\    
    5. 6 & $\mathrm{H}_{2} + \mathrm{e} \rightleftharpoons 2 \mathrm{H} + \mathrm{e}$ & $k_{5} = k_{5, \mathrm{H}}^{1-a} k_{5, \mathrm{L}}^{a} $ & \\
    & & $k_{5, \mathrm{H}} = 1.91 \times 10^{-9} T^{0.136} \mathrm{exp}\Bigl( - 53407.1/T \Bigr)$ & {\small \cite{Trevisan:2002}} \\
    & & $k_{5, \mathrm{L}} = 4.49 \times 10^{-9} T^{0.11} \mathrm{exp}\Bigl( - 101858/T \Bigr)$ & \\
    & & $a = \left( 1 + n_{\mathrm{H}} / n_{\mathrm{crit}} \right)^{-1}$ & \\        
    & & $n_{\mathrm{crit}} = \left[y(\mathrm{H})/n_{\mathrm{crit}}(\mathrm{H}) + 2y(\mathrm{H}_{2})/n_{\mathrm{crit}}(\mathrm{H}_{2}) 
       + y(\mathrm{He})/n_{\mathrm{crit}}(\mathrm{He})\right]^{-1}$ & \\            
    & & $\mathrm{log}~(n_{\mathrm{crit}}(\mathrm{H})) = 3 - 0.416 \  \mathrm{log}~(T/10^{4}) - 0.372 \left[ \mathrm{log}~(T/10^{4}) \right]^{2}$ & \\                
    & & $\mathrm{log}~(n_{\mathrm{crit}}(\mathrm{H}_{2})) = 4.845 - 1.3 \  \mathrm{log}~(T/10^{4}) + 1.62 \left[ \mathrm{log}~(T/10^{4}) \right]^{2}$ & \\                
    & & $\mathrm{log}~(n_{\mathrm{crit}}(\mathrm{He})) = 5.0792 \left[ 1 - 1.23 \times 10^{-5} (T - 2000) \right]$ & \\            
    7. 8 & $3 \mathrm{H} \rightleftharpoons \mathrm{H}_{2} + \mathrm{H}$ & $k_{7} = 7.7\times10^{-31}T^{-0.464}$ & {\small \cite{Glover:2008-3}} \\
    9, 10 & $2 \mathrm{H} + \mathrm{H}_{2} \rightleftharpoons 2 \mathrm{H}_{2}$ & $k_{9} = k_{7}/8$ & {\small \cite{Palla:1983}} \\
    11, 12 & $\mathrm{H}^{-} + \mathrm{H}^{+} \rightleftharpoons 2 \mathrm{H}$ & $k_{11} = 2.4 \times 10^{-6} T^{-0.5} \Bigl( 1.0 + T/20000 \Bigr)$ & {\small \cite{Croft:1999}} \\
    13, 14 & $\mathrm{H}^{+} + \mathrm{e} \rightleftharpoons \mathrm{H} + \gamma$ & 
       $k_{13} = 2.753 \times 10^{-14} \Bigl( 315614/T \Bigr)^{1.5} \Bigl[ 1.0 + \Bigl( 115188/T \Bigr)^{0.407} \Bigr]^{-2.242}$ & {\small \cite{Ferland:1992}} \\
    15, 16 & $\mathrm{H} + \mathrm{e} \rightleftharpoons \mathrm{H}^{-} + \gamma$ & 
       $k_{15} = \mathrm{dex}[-17.845 + 0.762 \mathrm{log}~T + 0.1523 (\mathrm{log}~T)^{2}$ & {\small \cite{Wishart:1979}} \\
    & & $~~~~~~ - 0.03274 (\mathrm{log}~T)^{3}]$ ~~~~~~~~ ($T < 6000~\mathrm{K}$) & \\
    & & $~~~ = \mathrm{dex}[-16.4199 + 0.1998 (\mathrm{log}~T)^{2} - 5.447 \times 10^{-3} (\mathrm{log}~T)^{4}$ & \\
    & & $~~~~~~ + 4.0415 \times 10^{-5} (\mathrm{log}~T)^{6}]$ ~ ($T > 6000~\mathrm{K}$) & \\
    17, 18 & $\mathrm{H}_{2} + \mathrm{He} \rightleftharpoons 2 \mathrm{H} + \mathrm{He}$ & $k_{17} = k_{17, \mathrm{H}}^{1-a} k_{17, \mathrm{L}}^{a} $ & \\    
    & & $k_{17, \mathrm{H}} =  \mathrm{dex}[-1.75~\mathrm{log}T -2.729 - 23474/T]$ & {\small \cite{Dove:1987}} \\
    & & $k_{17, \mathrm{L}} = \mathrm{dex}[3.801~\mathrm{log}T -27.029 - 29487/T]$ & \\
    19, 20 & $2 \mathrm{H} \rightleftharpoons \mathrm{H}^{+} + \mathrm{e} + \mathrm{H}$ & 
        $k_{19} = 1.2\times10^{-17}T^{1.2}~\mathrm{exp}\left( -\frac{157800}{T} \right)$ & {\small \cite{Lenzuni:1991}} \\         
    21 & $\mathrm{H}_{2} + \gamma_{\mathrm{ex}} \rightarrow \mathrm{H}_{2}^{\ast} \rightarrow 2\mathrm{H}$ & 
        $k_{21} = 1.4\times10^{9}J_{\mathrm{ex}}(h\nu=12.4~\mathrm{eV})f_{\mathrm{sh}}$  &  {\small \cite{Drain:1996}} \\
    & & $f_{\mathrm{sh}}=\mathrm{min}\left[ 1,~\left( \frac{N_{\mathrm{H}_{2}}}{10^{14}~\mathrm{cm}^{-2}} \right)^{-3/4} \right]$ & \\ 
    22 & $\mathrm{H}^{-} + \gamma_{\mathrm{ex}} \rightarrow \mathrm{H} + \mathrm{e}$ 
    & $k_{22} = \left[J_\mathrm{ex}(\nu)/B_{\nu}(T_{\mathrm{ex}})\right]k_{15}(T_{\mathrm{ex}})/K(T_{\mathrm{ex}})$ & \\ 
    & & $K(T_{\mathrm{ex}}) = \left[ \frac{n(\mathrm{H}^{-})}{n(\mathrm{H})n(\mathrm{e})} \right]^{\ast}$ & \\ \hline
  \end{tabular}
  }
  \\
 \begin{flushleft}
   Note.\textemdash The temperature $T_{\mathrm{e}}$ is in eV. The value of $N_{\mathrm{H}_2}$ is the column density of molecular hydrogen. 
 \end{flushleft} 
 \end{center}
\end{table*}

\section{Barotropic relation}
\label{App:b}

\begin{figure}
 \begin{center}
 \begin{tabular}{c} 
  {\includegraphics[width=0.95\columnwidth]{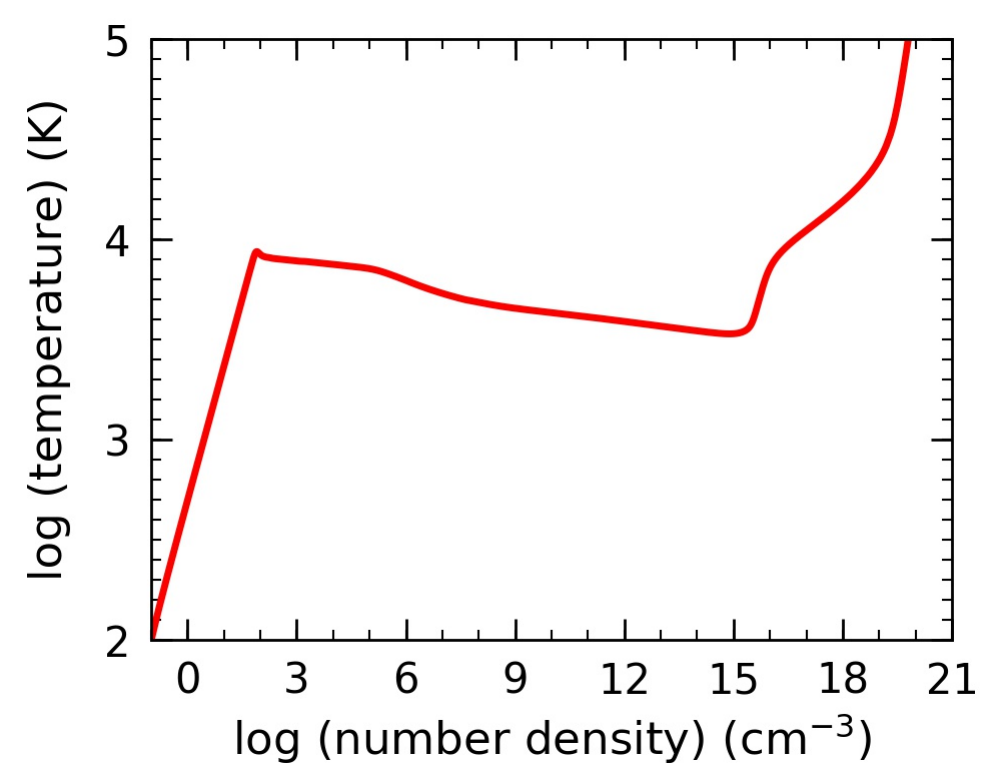}}
 \end{tabular}
 \caption{Temperature evolution in a one-zone calculation of gravitationally collapsing core. The horizontal axis is the number density and the vertical axis is the gas temperature.}
 \label{Fig:B1}
 \end{center}
\end{figure}

In Section~\ref{Sec:3-3}, we describe the results from the simulation with the barotropic temperature-density relation shown in Figure~\ref{Fig:B1}, for
comparison with the previous study \citep{Sakurai:2016}. 
To obtain this barotropic relation, we have carried out 
a one-zone calculation of the chemical and thermal evolution of a gravitationally collapsing core \citep{Omukai:2001},
using our thermal and chemical models.
Using the relation between the number density and temperature in Figure~\ref{Fig:B1}, with Equations~(\ref{Eq:5}) and (\ref{Eq:7}),
we obtain $P$ as a function of $\Sigma$.


\bsp	
\label{lastpage}
\end{document}